\documentclass[a4paper,11pt]{article}

\usepackage{amsmath,amsfonts,amssymb,caption,graphicx}
\usepackage{ushort}
\usepackage{makeidx}
\usepackage{float}
\usepackage{accents}
\usepackage{color}
\usepackage{framed}
\usepackage{versions}
\usepackage{xcolor}
\usepackage{emptypage}
\usepackage{mathbbol}
\usepackage{array}
\usepackage{multirow}

\def\hmath$#1${\texorpdfstring{{\rmfamily\textit{#1}}}{#1}}

\usepackage{tikz}
\usetikzlibrary{trees}

\usepackage[bookmarks]{hyperref}
\hypersetup{
    colorlinks=true,   	
    linkcolor=red,      
    citecolor = [rgb]{0 0.7 0},   	
    filecolor=magenta, 	
    urlcolor=blue
}
 
\input{epsf}

\newcommand{\weakly}{\mbox{$ \;\stackrel{\cal D}{\longrightarrow}\; $}}

\newcommand{\leqa}{\mbox{$ \;\stackrel{(a)}{\leq}\; $}}

\newcommand{\geqb}{\mbox{$ \;\stackrel{(b)}{\geq}\; $}}
\newcommand{\geqc}{\mbox{$ \;\stackrel{(c)}{\geq}\; $}}
\newcommand{\geqd}{\mbox{$ \;\stackrel{(d)}{\geq}\; $}}

\newcommand{\eqa}{\mbox{$ \;\stackrel{(a)}{=}\; $}}
\newcommand{\eqb}{\mbox{$ \;\stackrel{(b)}{=}\; $}}

\newcommand{\eqe}{\mbox{$ \;\stackrel{(e)}{=}\; $}}

\newcommand{\RL}{{\mathbb R}}

\newcommand{\IND}{{\mathbb I}}
\newcommand{\BBP}{{\mathbb P}}

\newcommand{\BBE}{{\mathbb E}}

\newcommand{\VAR}{\mbox{\rm Var}}

\newcommand{\TMAX}{T_{\rm MAX}}

\def\ba{\begin{align}}
\def\ea{\end{align}}
\def\ban{\begin{align*}}
\def\ean{\end{align*}}

\def\be{\begin{eqnarray}}
\def\ee{\end{eqnarray}}
\def\ben{\begin{eqnarray*}}
\def\een{\end{eqnarray*}}

\def\bqq{\begin{equation}}
\def\eqq{\end{equation}}
\def\bqqn{\begin{equation*}}
\def\eqqn{\end{equation*}}

\def\CTW{{\sf CTW}}

\def\BCT{{\sf BCT}}

\def\sq{$\Box$}

\def\qed{\ifmmode\sq\else{\unskip\nobreak\hfil
\penalty50\hskip1em\null\nobreak\hfil\sq
\parfillskip=0pt\finalhyphendemerits=0\endgraf}\fi\par\medbreak}

\newsavebox{\junk}
\savebox{\junk}[1.6mm]{\hbox{$|\!|\!|$}}

\def\limsup{\mathop{\rm lim\ sup}}
\def\liminf{\mathop{\rm lim\ inf}}

\def\til={{\widetilde =}}

\def\clA{{\cal A}}
\def\clB{{\cal B}}

\def\clJ{{\cal J}}

\def\clL{{\cal L}}
\def\clM{{\cal M}}

\def\clS{{\cal S}}
\def\clT{{\cal T}}

 \def\eq#1/{(\ref{#1})}

\newtheorem{theorem}{Theorem}[section]
\newtheorem{corollary}[theorem]{Corollary}
\newtheorem{proposition}[theorem]{Proposition}
\newtheorem{lemma}[theorem]{Lemma}

\def\eq#1/{(\ref{e:#1})}

\def\bdes{\begin{description}}
\def\edes{\end{description}}

\DeclareMathOperator*{\Moplus}{\text{\raisebox{0.25ex}%
	{\scalebox{0.8}{$\bigoplus$}}}}

\def\notes#1{}

\definecolor{mag}{rgb}{0.7,0,0.3}
\definecolor{dgreen}{rgb}{0.1,0.5,0.1}
\definecolor{dred}{rgb}{.8,0,0}
\definecolor{gray}{rgb}{.8,.8,.8}
\definecolor{brown}{rgb}{0.6451,0.3706,0.1745}

\begin{document}

\title{\vspace{-1.5cm}%
Context-Tree Weighting and Bayesian Context Trees:\\
Asymptotic and Non-Asymptotic Justifications}

\author
{
        Ioannis Kontoyiannis
    \thanks{Statistical Laboratory,
	Department of Pure Mathematics and Mathematical statistics,
        University of Cambridge,
	Wilberforce Road, Cambridge CB3 0WB, UK.
                Email: \texttt{\href{mailto:yiannis@maths.cam.ac.uk}%
			{yiannis@maths.cam.ac.uk}}.
        }
}

\date{\today}

\maketitle

\begin{abstract}
The Bayesian Context Trees (BCT) framework 
is a recently introduced, general collection 
of statistical and algorithmic tools for
modelling, analysis and inference 
with discrete-valued time series.
The foundation of this development is built
in part on some well-known information-theoretic
ideas and techniques, including Rissanen's tree sources 
and Willems et al.'s context-tree weighting algorithm.
This paper presents a collection of theoretical
results that provide 
mathematical justifications 
and further insight into the 
BCT modelling framework and the associated 
practical tools. It is shown that
the BCT prior predictive likelihood (the probability
of a time series of observations averaged over all
models and parameters) is both pointwise and 
minimax optimal, in agreement with the MDL principle 
and the BIC criterion. The posterior
distribution is shown to be asymptotically
consistent with probability one
(over both models and parameters),
and asymptotically Gaussian
(over the parameters).
And the posterior predictive 
distribution is also shown to be
asymptotically consistent with probability one.
\end{abstract}


\newpage

\section{Introduction}
\label{s:intro}

Let $x_1^n=(x_1,x_2,\ldots,x_n)$ be a time series of observations
with values in some finite alphabet $A$. Modelling $x_1^n$ as a realisation
of a random process, or {\em source}, $\{X_n\}$ is one of the
fundamental first
steps in developing algorithms for processing, compressing, 
transmitting, or performing any one of a number of statistical tasks
based on $x_1^n$.
The obvious first modelling choice 
for most discrete time series with significant temporal structure
is via higher-order Markov chains, but the severe limitations
of this approach were recognised early on. The class of
finite-memory chains is structurally poor (one only 
has control of the memory length) and the number of 
associated parameters
grows very quickly to prohibitively large numbers. For example,
to describe the distribution of a simple 5th order chain on a
4-letter alphabet already requires the specification
of over three thousand parameters.

The information-theoretic literature is one of the first places
where alternative approaches were developed for overcoming these
difficulties. The most notable such approach is probably the introduction
of variable-memory Markov models 
by Rissanen~\cite{rissanen:83b,rissanen:86},
in connection with his development of the celebrated 
Minimum Description Length 
(MDL) principle~\cite{rissanen:book,grunwald:book}.
These variable-memory models -- 
known under various names,
including {\em tree sources}, {\em FSMX sources},
and {\em finitely generated sources} -- have found
very successful application in numerous areas, 
including data compression~\cite{weinbergeretal:94,weinberger-et-al:95},
prediction~\cite{merhav-feder:98},
and model selection~\cite{buhlmann:04,bejerano:04},
among others.

Of particular importance in
the present setting is the 
context-tree weighting (\CTW) algorithm,
originally
introduced~\cite{willems-shtarkov-tjalkens:95,willems:98}
as a method for the compression of discrete time series
using such variable-memory Markov chain models.
At the same time as its importance for data compression
was being recognised, it was also  
gradually becoming clear that 
the \CTW~algorithm can be best understood 
once it is properly rooted on a firm statistical
foundation. Following a number of early relevant 
works~\cite{maxtree1,willems:96,%
ctw:00,ctw:02,nowbakht-willems:02},
progress along this direction culminated in the recent 
development~\cite{BCT-JRSSB:22}
of the {\em Bayesian Context Trees} (\BCT) 
framework. 

The \BCT~framework provides a general
Bayesian foundation within which the
\CTW~algorithm and its associated tools and
techniques can be explored in a systematic and 
principled fashion. Indeed, 
in \cite{BCT-JRSSB:22} it was shown
that the \CTW~algorithm along with 
two generalisations of the context-tree
maximising algorithm~\cite{ctw:00,ctw:02}
can be used
very effectively for Bayesian
inference with discrete time series,
in particular for model selection and prediction.
A Markov chain Monte Carlo sampler was also developed for the
posterior distribution over both models
and parameters, and a more efficient,
simple Monte Carlo sampler was introduced
in \cite{branch-isit:22,papag-K-pre:23},
where the application of the \BCT~framework
to estimation problems was explored further.
These ideas were extended to provide
effective methods for segmentation
and change-point detection of discrete
time series in~\cite{lungu-pap-K:22,lungu-arxiv:22}.
Finally, a perhaps somewhat surprising generalisation
of the \BCT~framework for {\em real-valued}
time series was introduced in~\cite{pap-K-trieste:22,BCTAR3:arxiv}.
Many of the algorithms described in these recent works
are implemented in the publicly available
{\sf R} package `\BCT'~\cite{BCT:Rv1.1}.
Similar models have been considered 
in~\cite{gotoh:98,goto:01}.

The main purpose of this paper is to 
derive a number of theoretical results
that can offer additional insight into the
performance of statistical and methodological
tools associated with
the \BCT~framework, and also provide
theoretical justification for their practical application.
Some of these results are in the 
form of classical information-theoretic or statistical
asymptotics, while others provide explicit
finite-blocklength bounds.

In Section~\ref{s:VLMC} we recall the \BCT~framework
and collect the definitions and basic properties
that will be used throughout the paper.
Section~\ref{s:theory} contains our main results,
and Section~\ref{s:proofs} contains their proofs.


\section{Preliminaries: Bayesian Context Trees} 
\label{s:VLMC}

Here we collect the necessary definitions, 
assumptions, and basic results that will be used throughout
the paper.  All the results of this section 
(except for the straightforward observations
in Proposition~\ref{prop:MLE}, proved in Section~\ref{s:proofs})
can be found, along with
more extensive discussion and details, 
in~\cite{BCT-JRSSB:22}.

\subsection{Variable-memory Markov chains}
\label{s:VLMC-1}

Let $\{X_n\}$ be a discrete random source, understood
as a random process taking values in a finite alphabet
$A$ of $m:=|A|\geq 2$ symbols; without loss of generality,
we assume throughout that $A=\{0,1,\ldots,m-1\}$.
The models we consider for $\{X_n\}$ are
variable-memory representations of Markov chains
with memory length no greater than some fixed $D\geq 0$.
These representations describe the conditional distribution
of $X_n$ given $X_{n-D}^{n-1}=(X_{n-D},X_{n-D+1},\ldots,X_{n-1})$ 
by specifying 
a {\em model} $T$ and an associated
{\em parameter vector} $\theta$;
throughout, we write $X_i^j$ for a vector
of random variables
$(X_i,X_{i+1},\ldots,X_j)$ and similarly
$x_i^j$ for a string
$(x_i,x_{i+1},\ldots,x_j) \in A^{j-i+1}$,
for $i\leq j$.

The {\em model} of $\{X_n\}$ is represented
by a tree $T$ from 
the class $\clT(D)$
of all proper $m$-ary trees
of depth no greater than $D$; $T$ is 
{\em proper} if all of the $m$ possible children
of every internal node of $T$ are also in $T$.
Viewing every node of $T$ as a {\em context}, 
namely, a string of no more than $D$ symbols from $A$,
and viewing $T$ as the collection of its leaves,
the {\em parameter vector} $\theta$ associated with $T$
is 
$\theta=\{\theta_s;s\in T\}$,
where each $\theta_s$ is 
a discrete probability vector, 
$$\theta_s=(\theta_s(0),\theta_s(1),\ldots,\theta_{s}(m-1)),$$
so that the $\theta_s(j)$ are nonnegative and sum to one,
$\sum_{j\in A}\theta_s(j)=1$,
for each $s\in T$.

\begin{figure}[ht!]
\centerline{\includegraphics[width=2.8in]{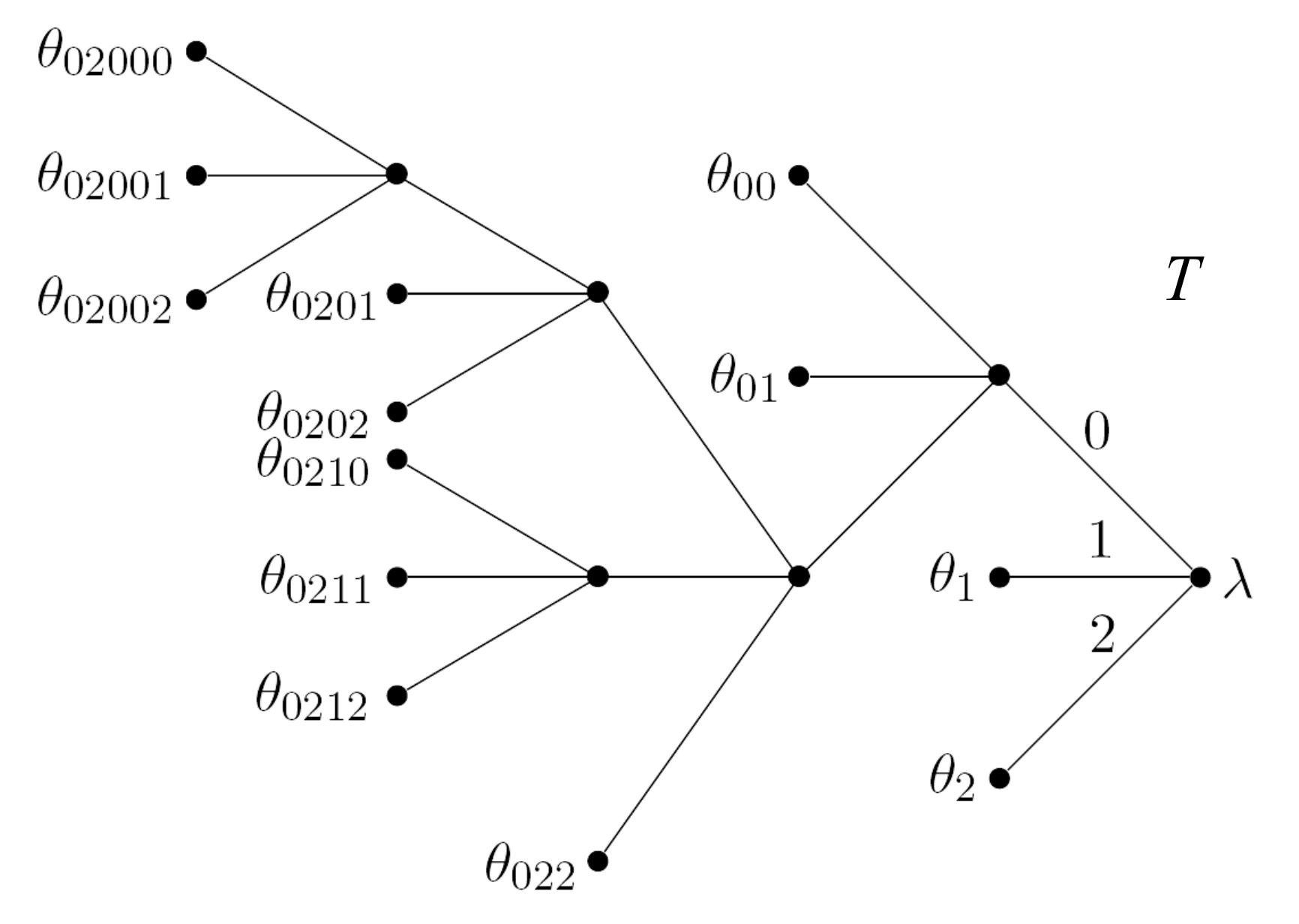}}
\caption{Example of a tree model $T$
for a ternary 5th order chain,
where the root $\lambda$ represents the empty string.
Also shown are the associated parameters $\theta_s$
for each leaf $s$ in $T$.}
\label{fig:running}
\end{figure}

A model $T\in\clT(D)$ together with an associated
parameter vector $\theta=\{\theta_s;s\in T\}$ specify
the conditional distribution of $X_n$ given $X_{n-D}^{n-1}$ 
as follows. Given $X_{n-D}^{n-1}=x_{n-D}^{n-1}$, let
$s$ denote the unique leaf of $T$ that is a suffix of $x_{n-D}^{n-1}$.
Then, for each $a\in A$,
$$\BBP(X_n=a|X_{n-D}^{n-1}=x_{n-D}^{n-1})=\theta_s(a).
$$
For example, consider the model $T$ of a 5th order chain
with alphabet $A=\{0,1,2\}$, represented by the tree shown in 
Figure~\ref{fig:running}. Then, conditional on 
$(X_{n-1},X_{n-2},X_{n-3},X_{n-4},X_{n-5})=(0,2,2,1,2)$,
the probability
that $X_n=1$ is,
$$P(1|02212):=\BBP(X_n=1|X_{n-5}^{n-1}=02212)=\theta_{022}(1),$$
where the relevant context now is $s=022$.
More generally, the probability of a block of observations
$x_1^n$ given an initial context $x_{-D+1}^0$ can be expressed,
via the Markov property, as,
\be
P(x_1^n|x_{-D+1}^0):=\BBP(X_1^n=x_1^n|X_{-D+1}^0=x_{-D+1}^0)
=
\prod_{i=1}^n P(x_i|x_{i-D}^{i-1})
=
\prod_{s\in T} \prod_{j\in A}
\theta_s(j)^{a_s(j)},
\label{eq:likelihood2}
\ee
where each element $a_s(j)$ of
the {\em count vector} $a_s=(a_s(0),a_s(1),\ldots,a_{s}(m-1))$
is,
\be
a_s(j)\;:=\;\mbox{\# times symbol $j\in A$ follows 
context $s$ in $x_1^n$}.
\label{eq:vector-a}
\ee

Given $T\in\clT(D)$, the family 
of chains with that model
can be parametrised by 
vectors $\phi=\{\phi_s;s\in T\}$ 
with each $\phi_s=(\phi_s(0),\ldots,\phi_s(m-2))$,
where the~$\phi$ belong to the set,
\be
\Omega(T,m):=\Big\{
\phi=\{\phi_s;s\in T\}
\in[0,1]^{|T|(m-1)}\;:\;
\sum_{j=0}^{m-2}\phi_s(j)\leq 1,
\;\mbox{for each}\;s\in T
\Big\},
\label{eq:omega}
\ee
which is a compact Euclidean subset of $\RL^{|T|(m-1)}$
with nonempty interior, and where $|T|$ denotes
the number of leaves of $T$. To each
$\phi\in\Omega(T,m)$ we naturally associate
a parameter vector $\theta=\theta(\phi)$ by letting, for each $s\in T$,
\be
\theta_s(j)=\phi_s(j),\;\mbox{for}\;0\leq j\leq m-2,
\;\mbox{and}\;\theta_s(m-1)=1-\sum_{0\leq j\leq m-2}\phi_s(j).
\label{eq:parametermap}
\ee

\subsection{Prior structure}
\label{s:prior}

{\bf Model prior. }
Given a fixed maximal depth $D\geq 0$ and an arbitrary 
$\beta\in(0,1)$, we define a prior distribution
on the collection $\clT(D)$ of models $T$ 
of depth no more than $D$, by,
\be
\pi(T):=\pi_D(T):=\pi_D(T;\beta):=\alpha^{|T|-1}\beta^{|T|-L_D(T)},
\label{eq:tree-prior}
\ee
where $\alpha:=(1-\beta)^{1/(m-1)}$,
and $L_D(T)$ denotes the number of leaves 
$T$ has at depth $D$. Clearly $\pi(T)$
penalises larger models by an exponential
amount, and the value of the hyperparameter
$\beta$ controls the degree of this 
penalisation. The fact that~(\ref{eq:tree-prior})
indeed defines a probability distribution and
the specific way in which larger models are penalised
are discussed in detail in~\cite{BCT-JRSSB:22}.

\medskip

\noindent
{\bf Prior on parameters. }
Given a model $T\in\clT(D)$, we define
the following prior distribution
on the parameter vectors $\theta=\{\theta_s;s\in T\}$:
We place an independent
Dirichlet distribution with parameters $(1/2,\ldots,1/2)$ 
(denoted Dir$(1/2,\ldots,1/2)$)
on each $\theta_s$ so that,
$\pi(\theta|T)=\prod_{s\in T}\pi(\theta_s),$
where,
\ben
\pi(\theta_s)
= \pi(\theta_s(0), \theta_s(1), \ldots, \theta_s({m-1}))
= \frac{\Gamma(m/2)}{\pi^{m/2}}\prod_{j=0}^{m-1}\theta_s(j)^{-\frac{1}{2}}
\propto \prod_{j=0}^{m-1}\theta_s(j)^{-\frac{1}{2}}.
\een

\noindent
{\bf Likelihood. }
Given a model $T\in\clT(D)$, the associated
parameter vector $\theta=\{\theta_s;s\in T\}$, 
and observations $x_1^n$ with initial context $x_{-D+1}^0$,
the {\em likelihood}
is given as in~(\ref{eq:likelihood2}),
\be
P(x_1^n|x_{-D+1}^0,\theta,T)
=
\prod_{s\in T} \prod_{j=0}^{m-1} \theta_s(j)^{a_s(j)}.
\label{eq:likelihood3}
\ee
In order to avoid
cumbersome notation, in what follows we often
write $x$ for the string $x_1^n$ and 
suppress the dependence on its initial 
context $x_{-D+1}^0$, so that, for example, 
we denote,
$
P(x|\theta,T)=P(x_1^n|x_{-D+1}^0,\theta,T).
$

\subsection{Marginal likelihood, posterior, 
and prior predictive likelihood}
\label{s:mml}

A useful property induced by the prior
specification above 
is that the parameters $\theta$ can
be integrated out, so that the 
{\em marginal likelihood} $P(x|T)$ can 
be expressed in closed form.

\begin{lemma}
\label{lem:integrate}
The marginal likelihood $P(x|T)$ 
of the observations
$x$ given a model $T$ is,
$$P(x|T)
=\int P(x,\theta|T)d\theta
=\int P(x|\theta,T)\pi(\theta|T)d\theta
=\prod_{s\in T}P_e(a_s),$$
where the count vectors 
$a_s$
are defined
in~{\em (\ref{eq:vector-a})}
and 
the {\em estimated probabilities} $P_e(a_s)$ are,
\be
P_{e}(a_s):= 
\frac{\prod_{j=0}^{m-1} 
[(1/2)(3/2)\cdots 
(a_s(j)- 1/2)]}
{(m/2)(m/2+1) 
\cdots (m/2+M_s-1)},
\label{eq:Pe}
\ee
where $M_s:=a_s(0)+\cdots+a_s(m-1)$, 
with the convention that any empty product
is equal to 1.
\end{lemma}

In terms of inference, the most important
quantity is the posterior distribution,
$$\pi(\theta,T|x)=\frac{P(x|\theta,T)\pi(\theta,T)}{P(x)},$$
where the main obstacle in its computation
is the {\em prior predictive likelihood} term,
\be
P(x):=P^*_D(x):=\sum_{T\in\clT(D)}\pi_D(T)P(x|T)
=\sum_{T\in\clT(D)}\int_\theta\pi_D(T)P(x|\theta,T)\pi(\theta|T)\,d\theta.
\label{eq:mml}
\ee
Nevertheless, the general version of the context-tree 
weighting (\CTW) algorithm can be used to 
efficiently compute 
the exact value of $P_D^*(x)$.

The \CTW\ first builds an $m$-ary tree, $\TMAX$, 
whose leaves are 
all the contexts $x_{i-D}^{i-1}$,
$i=1,2\ldots,n$,
that appear 
in the observations string $x_{-D+1}^n$,
together with any additional leaves required
so that $\TMAX$ is proper.
Then the estimated probabilities $P_{e,s}:=P_e(a_s)$ 
given by~(\ref{eq:Pe})
are computed 
at each node $s$ of $\TMAX$,
and finally
the {\em mixture} or {\em weighted probabilities}
are computed
at each node $s$ of $\TMAX$,
\be
P_{w,s}:=
\left\{
\begin{array}{ll}
P_{e,s}, &\;\;\mbox{if $s$ is a leaf,}\\
\beta P_{e,s}+(1-\beta)\prod_{j=0}^{m-1} P_{w,sj}, &\;\;\mbox{otherwise,}
\end{array}
\right.
\label{eq:Pw}
\ee
where
$sj$ denotes the concatenation
of context $s$ and symbol $j$, corresponding
to the $j$th child of node $s$.
The mixture probability $P_{w,\lambda}$ 
at the root $\lambda$ 
is exactly
$P^*_D(x).$

Similarly, the \BCT~algorithm efficiently
identifies the
maximum {\em a posteriori} probability (MAP)
context tree model $T_1^*$ that 
satisfies:
\be
\pi(T^*_1|x)=\max_{T\in\clT(D)}\pi(T|x).
\label{eq:thm-MAPT}
\ee
Proceeding as in the \CTW, after constructing
the tree $T_{\rm MAX}$ and computing the estimated
probabilities, now the {\em maximal probabilities}
are computed at each $s\in T_{\rm MAX}$:
\ben
P_{m,s}:=
\left\{
\begin{array}{ll}
P_{e,s}, &\;\;\mbox{if $s$ is a leaf at depth $D$,}\\
\beta, &\;\;\mbox{if $s$ is a leaf at depth $<D$,}\\
\max\Big\{\beta P_{e,s},\;(1-\beta)\prod_{j=0}^{m-1} P_{m,sj}, 
&\;\;\mbox{otherwise.}
\end{array}
\right.
\een
Then,
proceeding recursively from the root to its descendants, 
for each node $s$: If the above maximum can be achieved by 
the first term, then prune all its descendants from
$T_{\rm MAX}$; otherwise, repeat the same process at each of the 
children of node $s$. The resulting tree, $T_1^*$, 
after all nodes have been exhausted satisfies~(\ref{eq:thm-MAPT}).

Also we recall that
the {\em full conditional distribution} 
$\pi(\theta|x,T)$ of the parameters given
the model and observations,
is given by the following product of Dirichlet distributions:
\be
\pi(\theta|x,T)\sim
\prod_{s\in T}\mbox{Dir}
(a_s(0)+1/2,a_s(1)+1/2,\ldots,a_s(m-1)+1/2).
\label{eq:full-cond}
\ee

\subsection{Maximum likelihood and the posterior predictive distribution}
\label{s:MLE}

For a given data string $x$ and a
fixed depth $D$, the first steps of the \CTW~algorithm 
can be used to compute the maximum likelihood
estimates 
(MLEs)
for the model and parameters,
namely, the model $\hat{T}_{\rm MLE}$
and the associated parameters
$\hat{\theta}_{\rm MLE}:=\{\hat{\theta}_{s};s\in\hat{T}_{\rm MLE}\}$
that achieve:
\be
\hat{P}_{\rm MLE}(x):=
\max_{T\in\clT(D)}\sup_{\theta=\{\theta_s;s\in T\}}P(x|\theta,T).
\label{eq:MLE}
\ee


\begin{proposition}
\label{prop:MLE}
Let $x=x_{-D+1}^n$ be a given string of samples from $A$, let
$D\geq 1$,
and write $T_c(D)$ for the complete $m$-ary tree of depth $D$.
\begin{itemize}
\item[$(i)$]
The maximum likelihood  
over models $T\in\clT(D)$ in~{\em (\ref{eq:MLE})}
is equivalent to the classical maximum
likelihood among all Markov chains of memory
length $D$:
\ben
\hat{P}_{\rm MLE}(x)=\hat{P}_{\rm MLE}(x|T_c(D)):=
\sup_{\theta=\{\theta_s;s\in T_c(D)\}}P(x|\theta,T_c(D)).
\een
\vspace*{-0.1in}
\item[$(ii)$]
The maximum likelihood model
$\hat{T}_{\rm MLE}$ 
can always be taken to be $T_c(D)$.
The corresponding
maximum likelihood parameters 
$\hat{\theta}_{\rm MLE}=\{\hat{\theta}_s;s\in T_{\rm MLE}\}$
at each leaf $s$ are given by the empirical
frequencies $a_s/M_s$, where $a_s$ is 
given in {\em (\ref{eq:vector-a})}
and $M_s=\sum_j a_s(j)$, whenever
$a_s$ is not the all-zero vector.
If $a_s$ is zero, then $\hat{\theta}_{s}$ 
can be taken arbitrary.
\item[$(iii)$]
Equivalently, 
$\hat{T}_{\rm MLE}$ 
can be taken to be the tree
$\TMAX$ computed in the first step
of \CTW, with the parameter
vector $\hat{\theta}_{\rm MLE}=\{\hat{\theta}_s;s\in T_{\rm MAX}\}$
defined as before. 
\item[$(iv)$]
The actual maximum likelihood can be expressed as:
\vspace*{-0.05in}
\be
\hat{P}_{\rm MLE}(x)
=P(x_1^n|x_{-D+1}^0,\hat{\theta}_{\rm MLE},\hat{T}_{\rm MLE})
=\prod_{s\in \TMAX} \prod_{j=0}^{m-1} \hat{\theta}_s(j)^{a_s(j)}.
\label{eq:MLET}
\ee
\end{itemize}
\vspace*{-0.05in}
\end{proposition}

Finally we recall that, for the purposes of prediction,
standard Bayesian methodology dictates that 
the canonical rule for predicting the next 
observation $x_{n+1}$ given the past $x_1^n$,
is given by the {\em posterior predictive distribution},
\be
P^*_D(x_{n+1}|x_1^n)=\sum_T\int_\theta P(x_{n+1}|x_1^n,\theta,T)
\pi(\theta,T|x_1^n)\,d\theta.
\label{eq:defn}
\ee
where again, for simplicity, we suppressed the dependence
on the initial context $x_{-D+1}^0$.
A key observation is that,
using the \CTW, $P^*_D(x_{n+1}|x_1^n)$ can 
be computed exactly and
sequentially, as:
\be
P^*_D(x_{n+1}|x_1^n)=\frac{P^*_D(x_1^{n+1})}{P^*_D(x_1^n)}.
\label{eq:defn2}
\ee


\section{Main Results: Bounds and Asymptotics}
\label{s:theory}

The statistical tools provided by the \BCT\ framework have been 
found to provide efficient methods for very effective inference
in a variety of 
applications~\cite{BCT-JRSSB:22,papag-K-pre:23,lungu-arxiv:22,%
BCTAR3:arxiv,BCT:Rv1.1}.
In terms of the underlying theory,
the 
Bayesian perspective adopted in~\cite{BCT-JRSSB:22}
and this work 
is neither purely subjective, interpreting the prior
and posterior as subjective
descriptions of uncertainty pre- and post-data,
respectively,
nor purely objective,
treating the resulting methods as simple 
black-box procedures~\cite{chipman:01}.
For example, we think of the MAP model as the most 
accurate, data-driven representation of the 
regularities present in a given time series,
but we inform our analysis 
of the resulting inferential procedures
by simulation experiments on hypothetical models,
and by 
examining their
frequentist properties;
see, e.g.,~\cite[Chapter~6]{bernardo-smith:book}
or~\cite[Chapter~4]{gelman:book}
for broad discussions of the relationship
between the Bayesian and classical 
outlook.
This latter examination
is the main purpose of this paper.
Our main results, presented in this section,
provide classical 
asymptotic results as well as nonasymptotic bounds,
than can be viewed as partial justifications
of the \BCT\ framework.

A point of view which has had 
very significant influence in the
development of the ideas presented in
this work is Rissanen's celebrated
MDL principle.
As should become apparent from the  
form of the results in this section, there is 
also a strong connection with
Schwarz's Bayesian Information Criterion 
(BIC)~\cite{schwarz:78}, and its
familiar ``$(1/2)\log n$-per-degree-of-freedom'' 
log-likelihood penalty;
see~\cite{katz:81,mcculloch:91,kass:95,wasserman:00,csiszar-talata:06}
for extensive discussions of the role
of the BIC within Bayesian theory in general,
and its use in conjunction with Markov chain models.

Our main results are 
Theorems~\ref{thm:redundancy}--\ref{thm:shtarkov}
in Sections~\ref{s:MML}--\ref{s:shtarkov}.
As some of them are simple consequences of
known general results or generalisation
of previously established special cases, 
a detailed bibliographical discussion is
given in Section~\ref{s:history}. All proofs
are deferred to Section~\ref{s:proofs}.

\subsection{The prior predictive likelihood}
\label{s:MML}

The following three results show that the
logarithm of the prior predictive likelihood 
(cf.~(\ref{eq:mml}))
of {\em any} data string of length $n$,
is uniformly close to the log-likelihood of {\em every}
variable-memory chain, up to the best possible
penalty of order $\log n$.
Specifically, for 
every $x_1^n$
of arbitrary length $n$, 
any initial context $x_{-D+1}^0$,
and any model $T\in\clT(D)$
with parameters $\theta=\{\theta_s;s\in T\}$,
\be
\log P_D^*(x_1^n|x_{-D+1}^0)
\approx
\log P(x_1^n|x_{-D+1}^0,\theta,T)
	-
	\frac{|T|(m-1)}{2}
	\log n;
\label{eq:logn}
\ee
recall that $m$ denotes the alphabet size and
`$\log$' denotes the natural
logarithm throughout this work.
Moreover, this performance
is in a strong sense best possible.

The first result
states that the prior predictive likelihood
indeed achieves the performance announced
in~(\ref{eq:logn}), in a strong, nonasymptotic
sense.

\begin{theorem} 
\label{thm:redundancy}
For any variable-memory chain
with model $T\in\clT(D)$ 
and associated parameters $\theta=\{\theta_s;s\in T\}$, 
for any sequence $x_1^n$
of arbitrary length $n$,
and any initial context $x_{-D+1}^0$,
the prior predictive likelihood for any
$\beta$ satisfies,
\ben
\log P_D^*(x_1^n|x_{-D+1}^0)
\geq
\log P(x_1^n|x_{-D+1}^0,\theta,T)
	-
	\frac{|T|(m-1)}{2}
	\log n + C,
\een
where the constant $C=C(T,m,\beta)$ is 
independent of $n$ and of $\theta$, 
and can be taken,
$$
C=
C(T,m,\beta)
	=\frac{|T|(m-1)}{2}
	\log 
	|T|
	-|T|\log m
	+\log\pi_D(T;\beta),
\qquad\mbox{for }
n\geq e|T|.$$
\end{theorem}

The next result shows that
no other probability assignment 
can essentially outperform 
the prior predictive likelihood,
even on the average, and even
for a small fraction of
processes, as defined by their
relative volume in terms of 
the parametrisation in~(\ref{eq:omega}).
Theorem~\ref{thm:MDL}
is a simple consequence of a fundamental
result due to Rissanen~\cite{rissanen:84,rissanen:86b}.

\begin{theorem}
\label{thm:MDL}
Let $\{X_n\}$ denote an arbitrary
variable-memory chain with model
$T\in\clT(D)$ and associated
parameters $\theta=\{\theta_s;s\in T\}$, 
and suppose $\{Q_n\}$ is 
any consistent sequence 
of probability distributions
$Q_n$ on $A^n$, $n\geq 1$. Then, for every
$n$ and every $\epsilon>0$,
\be
\BBE_{\theta,T}[\log Q_n(X_1^n)]
\leq
\BBE_{\theta,T}[\log P(X_1^n|X_{-D+1}^0,\theta,T)]-
	(1-\epsilon)\frac{|T|(m-1)}{2}
	\log n,
\label{eq:MDL}
\ee
for all parameter vectors $\theta$,
except for those corresponding to
a subset $A_\epsilon(n)$ of $\Omega(T,m)$,
whose volume tends to zero as $n\to\infty$,
and where the expectation is taken 
with respect to the distribution
of the chain $\{X_n\}$.
\end{theorem}

The following result is analogous to that
of Theorem~\ref{thm:MDL}, except it states
the prior predictive likelihood will outperform
any other probability assignment $\{Q_n\}$
not just on the average but
on ``most'' sample strings $x_1^n$: The 
bound~(\ref{eq:MDL}) holds not only 
in expectation but in fact for most
$x_1^n$.  In order to state it precisely, 
we need the following notation. Given a model 
$T\in\clT(D)$
and an initial context $x_{-D+1}^0$, we say
that the strings $x_1^n$ and $y_1^n$
belong to the same {\em $T$-type},
if $x_{-D+1}^n$ and the concatenation
of $x_{-D+1}^0$ and $y_1^n$ induce
the same count vectors $a_s$ for all 
contexts $s\in T$; cf.~Lemma~\ref{lem:integrate}.
Then it is easy to see, e.g., that $A^n$
can be decomposed into polynomially many 
different such $T$-types, each of which
consists of exponentially many strings.
Let $M_n(T)$ denote the total number of $T$-types
of strings of length $n$.

Theorem~\ref{thm:WMF} is a consequence
of a general result due to Weinberger, Merhav 
and Feder~\cite{weinbergeretal:94}.

\begin{theorem}
\label{thm:WMF}
Let $\{Q_n\}$ be
any consistent sequence 
of probability distributions
$Q_n$ on $A^n$, $n\geq 1$,
and let $\epsilon>0$. 
Then, 
for every model $T\in\clT(D)$,
every parameter vector $\theta=\{\theta_s;s\in T\}$,
every sample size $n$, 
and every initial context $x_{-D+1}^0$,
we have,
$$\log Q_n(x_1^n)
\leq
\log P(x_1^n|x_{-D+1}^0,\theta,T)-
	(1-\epsilon)\frac{|T|(m-1)}{2}
	\log n,
$$
for all strings $x_1^n\in A^n$
except those in a set $B_n(\epsilon)\subset A^n$
which is asymptotically small in the sense that
the 
number $N_n(\epsilon,T)$ of $T$-types $\tau$
that contain a non-negligible part of $B_n(\epsilon)$,
i.e.,
$$\frac{|B_n(\epsilon)\cap\tau|}{|\tau|}>n^{-\epsilon/3},$$
is asymptotically negligible:
$$\frac{N_n(\epsilon,T)}{M_n(T)}\to 0,\qquad\mbox{as}\;n\to\infty.$$
\end{theorem}

Because the proofs of Theorems~\ref{thm:MDL}
and~\ref{thm:WMF} both depend of general 
information-theoretic results that are 
not proved here, in Section~\ref{s:shtarkov} 
we give a different upper bound which
can be proved
directly, without relying on any external results.

\subsection{The posterior predictive distribution}
\label{s:predictive}

Recall that a 
variable-memory chain $\{X_n\}$
with model $T\in\clT(D)$ 
and associated parameters $\theta=\{\theta_s;s\in T\}$
is {\em ergodic},
if the corresponding first order chain 
$\{Z_n:=X_{n-D+1}^n\;;\;n\geq 1\}$ 
is irreducible 
and aperiodic on a possibly strict subset
of $A^D$.
For an ergodic chain $\{X_n\}$
we write $\pi$ for its unique 
stationary distribution, as long as
this notation does not cause confusion 
with the similar notation used for the
posterior distribution $\pi(\theta,T|x)$.
In
order to avoid 
uninteresting technicalities, whenever we assume
that $\{X_n\}$ is ergodic, we 
implicitly also assume that its stationary
distribution $\pi$ gives strictly positive 
probability to all strings corresponding
to contexts $s$ in the model $T$.

As discussed in Section~\ref{s:MLE},
an important aspect of the \BCT~framework is that
the sequential nature of the \CTW~algorithm makes
it possible to efficiently compute the posterior 
predictive distribution, cf.~(\ref{eq:defn}) and~(\ref{eq:defn2}),
$$\BBP(X_{n+1}=j|x_{-D+1}^n)=P^*_D(j|x_{-D+1}^n)
=P^*_D(x_{-D+1}^nj)/P_D^*(x_{-D+1}^n),\qquad
j\in A.$$ 
Our next result,
Theorem~\ref{thm:pred}, states that
the posterior predictive 
distribution will in fact
converge to the true underlying distribution,
asymptotically with probability one.

\begin{theorem}
\label{thm:pred}
Suppose $\{X_n\}$ is an ergodic
variable-memory chain,
with model $T\in\clT(D)$ and associated 
parameters $\theta=\{\theta_s;s\in T\}$.
The posterior predictive distribution obtained 
from the prior predictive likelihood
with an arbitrary $\beta$
converges 
to the true conditional distribution
of the process $\{X_n\}$: 
For each $j\in A$, almost surely (a.s.) as $n\to\infty$:
$$
P^*_D(j|X_{-D+1}^n)-P(j|X_{-D+1}^n,\theta,T)
\to 0.
$$
\end{theorem}

\subsection{The posterior: Asymptotic consistency and normality}

Let $\{X_n\}$ be a variable-memory chain with model 
$T\in\clT(D)$. As discussed in~\cite{BCT-JRSSB:22}
the specific model $T$ that describes the chain is typically
not unique.
Of course, the main goal in model selection is to identify
the ``minimal'' model, that is, the smallest model that can
fully describe the distribution of the chain.

We call a model $T\in\clT(D)$ {\em minimal} with respect
to the 
parameter vector $\theta=\{\theta_s;s\in T\}$, 
if $T$ is either equal to $\Lambda:=\{\lambda\}$ 
or, if $T\neq\Lambda$,
then every $m$-tuple of leaves $\{sj\;;\;j=0,1,\ldots,m-1\}$
in $T$
contains at least two with non-identical parameters,
i.e., there are $j\neq j'$ such that $\theta_{sj}\neq\theta_{sj'}$.
It is easy to see 
that every $D$th order 
Markov chain $\{X_n\}$ has a unique minimal model 
$T^*\in\clT(D)$.
Throughout this section we will assume that $\{X_n\}$ 
is an ergodic $D$th order chain, with minimal model
$T^*\in\clT(D)$, associated 
parameters $\theta^*=\{\theta^*_s;s\in T^*\}$, 
unique stationary distribution denoted by $\pi$,
and an arbitrary initial context 
$x_{-D+1}^0$.

Given a sample $x=x_{-D+1}^n$,
a maximum {\em a posteriori} probability (MAP) model $T^*(n)$
within $\clT(D)$ is any $T\in\clT(D)$ which maximises
the posterior probability $\pi(T|x)$ over all 
$T\in\clT(D)$. Of course, the maximiser $T^*(n)$ need not 
always be unique.
Our next result says that, if the sample $x=x_{-D+1}^n$ is 
produced by an ergodic chain $\{X_n\}$ with minimal model $T^*\in\clT(D)$, 
then
the MAP model $T^*(n)$ is eventually unique and $T^*(n)=T^*$,
with probability one.

\begin{theorem}
\label{thm:modelC}
Let $\{X_n\}$ be an ergodic variable-memory
chain with minimal model $T^*\in\clT(D)$.
For any $\beta$, the MAP model $T^*(n)$ based on the random
sample $X_{-D+1}^n$ is eventually a.s.\ unique
and in fact:
$$T^*(n)=T^*,\qquad\mbox{eventually, a.s.}$$
\end{theorem}

\noindent
Using a different argument than that in the 
proof of Theorem~\ref{thm:modelC},
the following stronger consistency 
result can also be established.

\begin{theorem}
\label{thm:modelC2}
Under the assumptions of Theorem~\ref{thm:modelC},
the posterior distribution over models
eventually concentrates on $T^*$. For any $\beta$,
$$\pi(T^*|X_{-D+1}^n)\to 1,\qquad\mbox{a.s. as }n\to\infty.$$
\end{theorem}

Next we show that, as long as the true model
belongs to $\clT(D)$, the posterior distribution
on both the model and parameters eventually
a.s.\ concentrates around the true underlying 
values $(\theta^*,T^*)$. The proof of Theorem~\ref{thm:consistency}
uses Theorem~\ref{thm:modelC2}.

\begin{theorem}
\label{thm:consistency}
Let $\{X_n\}$ be an ergodic
variable-memory chain with minimal 
model $T^*\in\clT(D)$
and associated parameters $\theta^*=\{\theta^*_s;s\in T^*\}$;
let $\beta$ be arbitrary.
The posterior 
distribution $\pi(\theta,T|X_{-D+1}^n)$ 
asymptotically concentrates around
the true model and parameters,
i.e., 
\be\pi(\cdot|X_{-D+1}^n)\weakly \delta_{(\theta^*,T^*)},
\qquad\mbox{a.s., as }n\to\infty,
\label{eq:centering}
\ee
where $\weakly$ denotes weak convergence of 
probability measures, and $\delta_{(\theta^*,T^*)}$
is the unit mass at the point $(\theta^*,T^*)$.
\end{theorem}

Our next asymptotic result states that
the (appropriately centered and scaled) posterior 
distribution on the parameters is asymptotically normal. 
Recall that the density of the posterior 
can be decomposed as,
$$\pi(\theta,T|X_{-D+1}^n)=
\pi(T|X_{-D+1}^n)f_n(\theta|X_{-D+1}^n,T),$$
where $f_n(\theta|X_{-D+1}^n,T)$ is the 
full conditional density of the parameters
given in~(\ref{eq:full-cond}).
Since $\pi(T|X_{-D+1}^n)$ converges in distribution
to $\delta_{T^*}$, a.s., by Theorem~\ref{thm:modelC2}, 
we concentrate on the asymptotic
distribution $\pi(\theta|X_{-D+1}^n,T^*)$
of the parameters $\theta$ on $T^*$.
For its statement we will need the following 
notation. Given an ergodic chain with model $T^*$, 
stationary distribution $\pi$, and parameters $\theta^*$,
for each 
$s\in T^*$ let $J_s$ denote the 
$m\times m$ matrix,
\be
J_s=\frac{1}{\pi(s)}\left[\Theta^*_s-(\theta^*_s)^t(\theta^*_s)\right],
\label{eq:Js}
\ee
where $\Theta^*_s$ is the diagonal matrix
with entries $\theta^*_s(j)$, $j\in A$,
and $\theta^*_s$ is viewed as a row vector in $\RL^m$.
Then by $J$ we denote the $m|T^*|\times m|T^*|$ block-diagonal
matrix consisting of all $m\times m$ blocks $J_s$,
for $s\in T^*$,
\be
J=
\Moplus_{s\in T^*} J_s.
\label{eq:covariance}
\ee

\begin{theorem}
\label{thm:normality}
Let $\{X_n\}$ be an ergodic variable-memory Markov
chain with stationary distribution $\pi$, minimal 
model $T^*\in\clT(D)$ and associated parameters 
$\theta^*=\{\theta^*_s;s\in T^*\}$, 
with each $\theta_s^*(j)>0$; let $\beta$ be 
arbitrary. Suppose $\theta^{(n)}$
is distributed according to 
the posterior $\pi(\cdot|X_{-D+1}^n,T^*)$,
and let $\bar{\theta}^{(n)}$ denote its mean.
Then, as $n\to\infty$,
$$
\sqrt{n}\Big[\theta^{(n)}-\bar{\theta}^{(n)}\Big]
\weakly 
Z\sim N(0,J),\qquad\mbox{a.s.},$$
where $N(0,J)$ is the 
multivariate normal on $\RL^{m|T^*|}$ 
with zero mean and covariance matrix $J$ 
given in~(\ref{eq:covariance}); and also,
$\bar{\theta}^{(n)}\to\theta^*$ as $n\to\infty$:
$$
\bar{\theta}^{(n)}_s(j)=\BBE\left(\theta^{(n)}_s(j)\Big|X_{-D+1}^n,T^*\right)
\to\theta^*_s(j),
\qquad\mbox{a.s., for all $s\in T^*,\;j\in A$}.
$$
\end{theorem}

\subsection{An explicit minimax bound}
\label{s:shtarkov}

Finally we give a minimax version 
of Theorem~\ref{thm:MDL}. Theorem~\ref{thm:shtarkov}
gives
a more precise bound 
which is near-optimal
up to constant terms, not just the terms of order
$\log n$ as in Theorems~\ref{thm:MDL} and~\ref{thm:WMF}. 
On the other hand, it is 
a weaker version of Theorem~\ref{thm:WMF}:
It states that, for any probability assignment,
there is at least one variable-memory chain
and at least one string $x_1^n$ on which
it cannot outperform the prior predictive likelihood
asymptotically.

In order to make the comparison
between the upper and lower bounds
on $P_D^*$ more transparent,
before stating
Theorem~\ref{thm:shtarkov}
we give a simple corollary 
of Theorem~\ref{thm:redundancy}.
A close examination of its proof shows
that the following more detailed
bound is actually established there.

\begin{corollary}
For any variable-memory chain
with model $T\in\clT(D)$ 
and associated parameters $\theta=\{\theta_s;s\in T\}$, 
for any sequence $x_1^n$
of arbitrary length $n$,
and any initial context $x_{-D+1}^0$,
let $\clL(Q_n,x_1^n|\theta,T)$ denote the log-likelihood
ratio between an arbitrary 
probability distribution $Q_n$ on~$A^n$
and the true underlying
distribution of the chain:
$$\clL(Q_n,x_1^n|\theta,T):=
\log Q_n(x_1^n)-\log P(x_1^n|x_{-D+1}^0,\theta,T).$$
Then, for any 
$\beta$, the prior predictive likelihood
$P_D^*$ achieves,
\be
\clL(P_D^*,x_1^n|\theta,T)
\geq
-\left[\sum_{s\in T:M_s\neq 0}\left(\frac{m-1}{2}\log M_s+\log m\right)
	-\log\pi_D(T;\beta)\right],
\label{eq:corollary}
\ee
where $M_s$ are the sums of the count vectors $a_s$
as in Lemma~\ref{lem:integrate}.
\end{corollary}

Our final result 
shows that the performance achieved by 
the prior predictive likelihood $P_D^*$,
as described in~(\ref{eq:corollary}), cannot 
be improved upon by any sequence 
of probability distributions:
The log-likelihood of any
such choice will asymptotically be no better
than that of $P^*_D$ on at least one realisation 
produced by some variable-memory chain, 
up to a constant term that depends 
only on the alphabet size and the maximal memory
length $D$.

\begin{theorem}
\label{thm:shtarkov}
Suppose $\{Q_n\}$ is 
any (not necessarily consistent)
sequence of probability distributions
$Q_n$ on $A^n$, $n\geq 1$. Then, for any $\beta$,
any $D\geq 1$, and any initial context $x_{-D+1}^0$,
\begin{align*}
&\limsup_{n\to\infty}\min_{T\in\clT(D)}\inf_{\phi\in\Omega(T,m)}
\min_{x_1^n\in S_n}\\
&\left[
\clL(Q_n,x_1^n|\theta(\phi),T)
+
	\sum_{s\in T:M_s\neq 0}\left(
	\frac{m-1}{2}\log M_s
	+\log\Big(\frac{\sqrt{2\pi}}{
	2^{m/2}
	\Gamma(m/2)}\Big)\right)
	-\log\pi_D(T;\beta)
\right]\leq 0,
\end{align*}
where 
$\theta(\phi)$ for $\phi\in\Omega(T,M)$ refers
to the parametrisation of 
variable-memory chains given in~(\ref{eq:parametermap}),
and $S_n=S_n(\theta,T,x_{-D+1}^0)$ denotes
the collection of all strings $x_1^n\in A^n$
that have positive probability under 
$(\theta,T)$: $P(x_1^n|x_{-D+1}^0,\theta,T)>0$.
\end{theorem}

Observe that the difference between the
log-likelihood achieved by 
$P_D^*$ in~(\ref{eq:corollary}) and the
minimax optimality bound in Theorem~\ref{thm:shtarkov}
is small, it depends only on the alphabet size $m$, 
and it corresponds to a constant penalty $\Delta_m$
per model leaf, where $\Delta_m$ is simply,
$$\Delta_n=\log m 
-\log\left(\frac{\sqrt{2\pi}}{2^{m/2}\Gamma(m/2)}\right).$$

The bound in Theorem~\ref{thm:shtarkov} can be viewed
as a generalisation of Shtarkov's minimax
redundancy theorem in~\cite{shtarkov:87}.
Sharp minimax results in the same spirit,
for both i.i.d.\ and more general Markov processes,
are given in~\cite{xie-barron:00,jac-spa:04,barron-et-al:13}.
And for any specific model $T\in\clT(D)$, precise asymptotics
for the `minimax regret,'
$$\min_{Q_n}\sup_{\phi\in\Omega(T,m)}\max_{x_1^n\in S_n}
\big[-\clL(Q_n,x_1^n|\theta(\phi),T)\big],$$
are developed in~\cite{barron:14}.

\subsection{History and bibliographical remarks}
\label{s:history}

The lower bound on the prior predictive
likelihood in Theorem~\ref{thm:redundancy}, 
although 
essentially implicit in the existing literature,
is new in the form presented here;
it was established for the special case
of binary data ($m=2$) and $\beta=1/2$ 
in~\cite{ctw-4:93, willems-shtarkov-tjalkens:95},
and a version for general $m$ but only 
a specific value of $\beta=\beta(m)$ was given 
in~\cite{ctw-2:93,ctw-3:93,ctw-multi:94}.
The corresponding lower bound in expectation
given in Theorem~\ref{thm:MDL} is a straightforward
corollary of Rissanen's celebrated results
in~\cite{rissanen:84,rissanen:86b}.
Similarly, the lower bound in Theorem~\ref{thm:WMF}
follows from the general results in~\cite{weinbergeretal:94}.
A weaker, asymptotic version of the bounds in 
Theorems~\ref{thm:redundancy}--\ref{thm:WMF}
is established, under stronger assumptions in~\cite{gotoh:98}. 
The asymptotic consistency of the posterior predictive
distribution stated in Theorem~\ref{thm:pred} 
was first given in the special case $\beta=1/2$
in~\cite[Lemma~2]{weissman-et-al:di}.
Versions of 
the consistency result in 
Theorem~\ref{thm:modelC}
and the minimax lower bound in Theorem~\ref{thm:shtarkov} 
for binary data and $\beta=1/2$ are given 
in~\cite{willems-shtarkov-tjalkens:pre,ctw-4:93},
and similar techniques were used to establish
the eventual-a.s.\ consistency of the BIC 
estimator for context trees in~\cite{garivier:06};
the general results as stated and proved here are new.
Theorems~\ref{thm:modelC2},~\ref{thm:consistency}
and~\ref{thm:normality} are new.
An asymptotic relation similar to Theorem~\ref{thm:modelC},
and conditions for an implicit version of 
Theorem~\ref{thm:normality}, under additional assumptions,
are discussed in~\cite{gotoh:98}. 
A stronger version of Theorem~\ref{thm:modelC2}, including
a rate of convergence, but under slightly stronger assumptions,
was recently shown in~\cite{papag-K-pre:23}.
Earlier versions of Theorems~\ref{thm:modelC2}
and~\ref{thm:normality} are described, in terms 
of data compression, in~\cite{goto:01}.


\section{Proofs}
\label{s:proofs}

\noindent
We note for later use the following simple property
of the prior $\pi_D(T)$. 

\begin{lemma}
\label{lem:unions}
If $T\in\clT(D)$, $t\in T$ is at depth $d<D$, 
$S\in \clT(D-d)$ is nonempty, and $T\cup S$ consists 
of the tree $T$ with $S$ added as a subtree rooted
at $t$, then,
$$\pi_D(T\cup S) = \beta^{-1}\pi_D(T)\pi_{D-d}(S).$$
\end{lemma}

\noindent
{\em Proof. }
The result immediately follows from the simple
observations that
$|T\cup S|=|T|+|S|-1$ and
$L_D(T\cup S)=L_D(T)+L_{D-d}(S)$,
together with the definition of the prior.
\qed

\noindent
{\em Proof of Proposition~\ref{prop:MLE}. }
For the equivalence in~$(i)$, observe that,
in the present setting,
a Markov chain with memory length $D$
is equivalent to a variable-memory
chain with tree model corresponding 
to the complete tree $T_c(D)$. On the other
hand, a variable-memory chain with model given
by some tree $T$ can be represented as a 
chain with memory length $D$
by extending~$T$ to the 
complete tree and assigning to each new
leaf the same parameter vector as its
most recent ancestor in $T$. 

The above argument also shows that we
can always take $\hat{T}_{\rm MLE}$
to be the complete tree $T_c(D)$.
Then, to maximise with respect 
to the parameters $\theta$, 
recall from~(\ref{eq:likelihood3})
that the log-likelihood is,
\ben
\log P(x|\theta,T_c(D))
=
\sum_{s}
\sum_{j=0}^{m-1} a_s(j)\log\theta_s(j)
=
\sum_{s} M_s
\sum_{j=0}^{m-1} \frac{a_s(j)}{M_s}\log\theta_s(j),
\een
where we sum over all leaves $s$ of the
complete tree for which $a_s$ is not
the all-zero count vector,
and $M_s=\sum_ja_s(j)$. Writing 
$\hat{p}_s(j)=a_s(j)/M_s$, this becomes,
\ben
\log P(x|\theta,T_c(D))
&=&
\sum_{s} M_s\left\{
\sum_{j=0}^{m-1} \hat{p}_s(j)\log\left(\frac{\theta_s(j)}{\hat{p}_s(j)}\right)
+\sum_{j=0}^{m-1} \hat{p}_s(j)\log\hat{p}_s(j)
\right\}\\
&=&
-\sum_{s} M_s D(\hat{p}_s\|\theta_s)
-\sum_{s} M_s H(\hat{p}_s),
\een
where $H(p)$ and $D(p\|q)$
denote the entropy and relative entropy,
respectively, in nats.
Therefore, the likelihood is maximised by making
each divergence above equal to zero, i.e.,
by taking each $\theta_s=\hat{p}_s$ for leaves
$s$ with nonzero count vectors $a_s$. This proves~$(ii)$.

The fact claimed in~$(iii)$, that we can take 
$\hat{T}_{\rm MLE}=\TMAX$, is an immediate consequence
of the above computation, combined with 
the observation that the only leaves $s$ of 
$\TMAX$ at depth strictly smaller than $D$
have all-zero count vectors $a_s$.
Finally, the result of part~$(iii)$ together
with the simple expression for the likelihood
in~(\ref{eq:likelihood3}) show that the expression
in~(\ref{eq:MLET}) indeed computes the required
maximised likelihood.
\qed

\noindent
{\em Proof of Theorem~\ref{thm:redundancy}. }
For the sake of clarity, we adopt 
the simpler notation of 
Sections~\ref{s:prior} and~\ref{s:mml}.
Our starting point is the 
following pair of bounds on the
probabilities $P_e(a)$;
they follow from rather involved
but elementary computations and 
are stated here without proof. The
results are implicit 
in~\cite{krichevsky-trofimov:81,ctw-3:93,ctw-multi:94},
and slightly different proofs are given 
in~\cite{catoni:04}; see also~\cite{xie-barron:00}
for more detailed bounds.

\begin{lemma}
\label{lem:KT}
For any count vector $a=(a(0),a(1),\ldots,a(m-1))$
and $M=a(0)+\cdots+a(m-1)$, 
the probabilities $P_e(a)$ defined
in Lemma~\ref{lem:integrate} satisfy:
\be
\log P_e(a)&\geq&
\sum_{j=0}^{m-1}a(j)\log\frac{a(j)}{M}
-\frac{m-1}{2}\log M-\log m;
\label{eq:PeLB}\\
\log P_e(a)&\leq&
\sum_{j=0}^{m-1}a(j)\log\frac{a(j)}{M}
-\frac{m-1}{2}\log\frac{M}{2\pi}-\log\frac{\pi^{m/2}}{\Gamma(m/2)}.
\label{eq:PeUB}
\ee
\end{lemma}

We can now bound the marginal likelihoods
$P(x|T)$ for any 
$T\in\clT(D)$ and any parameter vector $\theta=\{\theta_s;s\in T\}$
as,
\ben
\log P(x|T)
&\eqa&
	\sum_{s\in T}\log P_e(a_s)\\
&\geqb&
	\sum_{s\in T:M_s\neq 0}\left\{
	\sum_{j=0}^{m-1}\left[a_s(j)\log\frac{a_s(j)}{M_s}\right]
	-\frac{m-1}{2}\log M_s-\log m\right\}\\
&=&
	\sum_{s\in T:M_s\neq 0}
	\log\left(
	\prod_{j=0}^{m-1}
	\Big(\frac{a_s(j)}{M_s}\Big)^{
	a_s(j)}\right)
	-\sum_{s\in T:M_s\neq 0}\left\{
	\frac{m-1}{2}\log M_s+\log m\right\}\\
&\geqc&
	\sum_{s\in T}
	\log\left(
	\prod_{j=0}^{m-1}
	\theta_s(j)^{
	a_s(j)}\right)
	-\frac{m-1}{2}
	\sum_{s\in T:M_s\neq 0}
	\log M_s-|T_n|\log m,
\een
where $(a)$ follows by Lemma~\ref{lem:integrate},
$(b)$ follows from~(\ref{eq:PeLB}) in Lemma~\ref{lem:KT},
$(c)$ follows form the fact that, 
as in the proof of Proposition~\ref{prop:MLE},
the empirical frequencies maximise the likelihood
over all parameter choices,
and $|T_n|$ denotes the number 
of $s\in T$ for which $M_s\neq 0$.
Therefore,
\ben
\log P(x|T)
&\geq&
	\log P(x|\theta,T)
	-
	\frac{|T_n|(m-1)}{2}
	\sum_{s\in T:M_s\neq 0}\frac{1}{|T_n|}
	\log M_s
	-|T_n|\log m\\
&\geqd&
	\log P(x|\theta,T)
	-
	\frac{|T_n|(m-1)}{2}
	\log \left(
	\sum_{s\in T:M_s\neq 0}\frac{1}{|T_n|}
	M_s\right)
	-|T_n|\log m\\
&\eqe&
	\log P(x|\theta,T)
	-
	\frac{|T_n|(m-1)}{2}
	\log \left(
	\frac{n}{|T_n|}
	\right)
	-|T_n|\log m,
\een
where $(d)$ follows Jensen's inequality 
and the concavity of the logarithm,
and $(e)$ follows from the observation
that $\sum_{s\in T:M_s\neq 0}M_s=n$.

Using the above inequality,
the prior predictive likelihood
$P_D^*(x)
=\sum_T\pi_D(T;\beta)P(x|T)$
can now be trivially bounded as,
\begin{align*}
\log P_D^*(x)
&\geq
\log\big(\pi_D(T;\beta)
P(x|T) 
\big)\\
&\geq
\log P(x|\theta,T)-
	\frac{|T_n|(m-1)}{2}
	\log \Big(
	\frac{n}{|T_n|}
	\Big)
	-|T|\log m
	+\log\pi_D(T;\beta)\\
&\geq
\log P(x|\theta,T)-
	\frac{|T|(m-1)}{2}
	\log \Big(
	\frac{n}{|T|}
	\Big)
	-|T|\log m
	+\log\pi_D(T;\beta),
\end{align*}
where the last inequality holds only for $n\geq e|T|$,
as required.
\qed

\noindent
{\em Proof of Theorem~\ref{thm:MDL}. }
The result is a more or less immediate
consequence of~\cite[Theorem~1]{rissanen:86b},
once we verify its assumptions. Recall the
parametrisation of all chains with model $T$
given in~(\ref{eq:omega}).
For any given
chain with model $T$, its parameters
$\phi=\{\phi_s\;;\;s\in T\}$ can be 
estimated from 
a sample $x_{-D+1}^n$ by the maximum
likelihood estimates,
$\hat{\phi}_s(j)=a_s(j)/M_s$,
for $s\in T$, $j=0,1,\ldots,m-2$.
Then the collection of estimates $\hat{\phi}
=\{\hat{\phi}_s\;;\;s\in T\}$ is asymptotically
normal, as established, e.g., by Billingsley
in~\cite{billingsley:markov,billingsley:61}.
The final condition requiring that, for
the class of processes considered here, 
$$\sum_{n\geq 1}
\BBP\{\sqrt{n}\|\hat{\phi}-\phi\|_1\geq\log n\}<\infty,$$
where $\|\cdot\|_1$ denotes
the $L^1$ norm,
is verified in~\cite{rissanen:86}.
\qed

\noindent
{\em Proof of Theorem~\ref{thm:WMF}. }
The result of the theorem is simply
a special case of~\cite[Theorem~1]{weinbergeretal:94}.
The probability assignment $\{Q_n\}$
corresponds to scheme $\clM$, 
and the pair $(\theta,T)$ corresponds
to a finite-state machine $F$ there.
A simple computation like that 
performed in the proof of Proposition~\ref{prop:MLE}
shows that their conditional entropy
$\hat{H}(x_1^n|F)$ is exactly the
negative of the normalised
maximum log-likelihood,
$$-\frac{1}{n}\log\hat{P}_{\rm MLE}(x_1^n|x_{-D+1}^0,T),$$
in the notation of Proposition~\ref{prop:MLE}.
And noting that, by definition,
$$-\frac{1}{n}\log\hat{P}_{\rm MLE}(x_1^n|x_{-D+1}^0,T)
\leq
-\frac{1}{n}\log P(x_1^n|x_{-D+1}^0,\theta,T),$$
Theorem~\ref{thm:WMF} is an immediate corollary 
of~\cite[Theorem~1]{weinbergeretal:94}.
\qed

\noindent
{\em Proof of Theorem~\ref{thm:pred}. }
We will need the
following asymptotic result on the ratios
of the probabilities $P_e(a)$ to the mixture probabilities
computed by \CTW:

\begin{lemma}
Under the assumptions of Theorem~\ref{thm:pred},
given a sample $x_{-D+1}^n$
as in \CTW, let
$P_{e,s,n}$ and $P_{w,s,n}$ denote the
probabilities 
at each node $s$ of $T$, as 
defined in~{\em (\ref{eq:Pe})} and~{\em (\ref{eq:Pw})},
respectively. Then, for any internal node $s$
of $T$, 
almost surely (a.s.), 
as $n\to\infty$:
$$\zeta_{s,n}:=\frac{P_{e,s,n}}{\prod_{j=0}^{m-1}P_{w,sj,n}}
\to 0.$$
\label{lem:gamma}
\end{lemma}

\noindent
The result of the lemma 
was first stated for the special
case $\beta=1/2$ in~\cite{C-K-Verdu:06b,weissman-et-al:di};
the proof of the general case 
follows along similar lines.

\medskip

\noindent
{\em Proof outline. }
From the definitions of 
$\zeta_{s,n}$, $P_{e,s,n}$ and $P_{w,sj,n}$, we have,
\ben
\frac{\zeta_{s,n}}{\beta\zeta_{s,n}+(1-\beta)}
&=&
	\frac{P_{e,s,n}}
	{\beta P_{e,s,n}+(1-\beta)\prod_{j=0}^{m-1}P_{w,sj,n}}\\
&\leq&
	\frac{P_{e,s,n}}
	{(1-\beta)\prod_{j=0}^{m-1}P_{w,sj,n}}\\
&\leq&
	\frac{P_{e,s,n}}
	{(1-\beta)^{m+1}\prod_{j=0}^{m-1}P_{e,sj,n}}\\
&\leq&
	(1-\beta)^{-m-1}
	\exp\left\{M_s\left[
	\frac{1}{M_s}\log P_{e,s,n}
	-\frac{1}{M_s}
	\log \prod_{j=0}^{m-1}P_{e,sj,n}
	\right]
	\right\}.
\een
Therefore, to prove the claimed result it suffices
to show that the exponential in the above right-hand-side
converges to zero a.s., because that would imply that
$\zeta_{s,n}/[\beta\zeta_{s,n}+(1-\beta)]\to 0$
a.s.,
which would in turn prove the lemma. But that is exactly
the content of the last part of the proof 
of~\cite[Lemma~12]{weissman-et-al:di}, where we
observe that the stationarity assumption can
be removed, in view of the classical ergodic
theorem for Markov chains~\cite{chung:book,meyn-tweedie:book2}.
\qed

In order to compute the ratio of the 
two prior predictive likelihoods that
defines the posterior predictive distribution, 
we first recall the sequential 
updating procedure described
in Section~3.5 of~\cite{BCT-JRSSB:22}:
In the notation of Lemma~\ref{lem:gamma}, 
having computed the count vectors
$a_{s,n}$, the probabilities
$P_{e,s,n}$ and the mixture probabilities
$P_{w,s,n}$ at each node of the tree
$\TMAX$, based on $x_{-D+1}^n$,
and given an additional sample $x_{n+1}=j$,
let $s_D,s_{D-1},\ldots,s_0$ denote the contexts
of length \mbox{$D,D-1,\ldots,0$}, respectively,
immediately preceding $x_{n+1}=j$.
For the computation of 
$P_{w,\lambda,n+1}=P^*_D((x_1,\ldots,x_n,j)|x_{-D+1}^0)$:
\begin{itemize}
\vspace*{-0.05in}
\item
At each 
of the nodes $s_D,s_{D-1},\ldots,s_0$,
the count vectors are updated as
$a_{s,n+1}(j)=a_{s,n}(j)+1$
and $M_{s,n+1}=M_{s,n}+1$;
at all other nodes $s$, 
$a_{s,n+1}(j)=a_{s,n}(j)$
and $M_{s,n+1}=M_{s,n}$.
\vspace*{-0.05in}
\item
At each 
of the nodes $s_D,s_{D-1},\ldots,s_0$,
the probabilities $P_e$ are updated
by
$P_{e,s,n+1}
=P_{e,s,n}P_{e,s,n+1|n}$, where,
$$P_{e,s,n+1|n}
:= \frac{a_{s,n+1}(j)-1/2}{m/2+M_{s,n+1}-1},$$
and we let
$P_{e,s,n+1} =P_{e,s,n}$, at all other nodes $s$.
\vspace*{-0.05in}
\item
Finally the mixture probabilities are
updated:
For $s=s_D$, which is necessarily 
a leaf, let $P_{w,s,n+1}=P_{e,s,n+1}$,
so that,
$$P_{w,s,n+1|n}
:=\frac{P_{w,s,n+1}}{P_{w,s,n}}
=\frac{P_{e,s,n+1}}{P_{e,s,n}}
= P_{e,s,n+1|n}.
$$
For the contexts $s=s_{D-1},\ldots,s_0$
which correspond to internal nodes, let,
$$P_{w,s,n+1}=\beta P_{e,s,n+1} + (1-\beta)\prod_{\ell=0}^{m-1}
	P_{w,s\ell,n+1},$$
as in~(\ref{eq:Pw}), so that,
\be
P_{w,s,n+1|n}
:=
	\frac{P_{w,s,n+1}}{P_{w,s,n}}
=
	\frac{
	\beta P_{e,s,n+1} + (1-\beta)\prod_{\ell=0}^{m-1}
	P_{w,s\ell,n+1}}
	{P_{w,s,n}}.
	\label{eq:update}
\ee
And for all other nodes $s$, we keep
$P_{w,s,n+1}=P_{w,s,n}$.
\end{itemize}
\vspace*{-0.05in}
Now, let $s=s_t$ be one of the contexts
$s_{D-1},\ldots,s_0$, and write $s_{t+1}$
as the concatenation $s_{t+1}=sa$ for some
$a\in A$. Then 
$P_{w,s\ell,n+1}=P_{w,s\ell,n}$ for all
$\ell\neq a$, and, therefore,~(\ref{eq:update})
gives,
\be
P_{w,s,n+1|n}
&=&
	\frac{
	\beta P_{e,s,n}P_{e,s,n+1|n} + 
	(1-\beta)P_{w,sa,n+1|n}\prod_{\ell=0}^{m-1} P_{w,s\ell,n}}
	{P_{w,s,n}}
	\nonumber\\
&=&
	\left(\frac{
	\beta\zeta_{s,n}
	}
	{(1-\beta)+\beta\zeta_{s,n}}\right)
	P_{e,s,n+1|n} + 
	\left(
	\frac{
	1-\beta
	}
	{(1-\beta)+\beta\zeta_{s,n}}
	\right)
	P_{w,sa,n+1|n},
	\label{eq:recursion}
\ee
by the definition of $\zeta_{s,n}$ in Lemma~\ref{lem:gamma}.

Now, in order to estimate the
posterior predictive probability,
$$P^*_D(j|x_{-D+1}^n)
=\frac{P^*_D((x_1,\ldots,x_n,j)|x_{-D+1}^0)}{P^*_D(x_1^n|x_{-D+1}^0)}
=\frac{P_{w,\lambda,n+1}}{P_{w,\lambda,n}}
=P_{w,\lambda,n+1|n},
$$
we observe that this can be done
recursively, starting from 
the leaf $s_D$ where
$P_{w,s_D,n+1|n}
= P_{e,s_D,n+1|n},$
and then proceeding through
$s_{D-1},\ldots,s_1$ all the
way to the root $\lambda=s_0$
via successive
applications of~(\ref{eq:recursion}),
until 
$P_{w,\lambda,n+1|n}$ is expressed
as a linear combination of the
conditional probabilities,
$P_{e,s_t,n+1|n},$ $t=D,D-1,\ldots,0$. 
It is easy to see
the coefficient of 
$P_{e,s_D,n+1|n}$ in this
linear combination is,
\be
\prod_{t=D-1}^0
	\frac{
	1-\beta
	}
	{(1-\beta)+\beta\zeta_{s_t,n}},
\label{eq:Cleaf}
\ee
while for 
all internal contexts
$s_T,$ $T=D-1,\ldots,0$, 
the coefficient of 
$P_{e,s_T,n+1|n}$ is,
\be
	\frac{
	\beta\zeta_{s_T,t}
	}
	{(1-\beta)+\beta\zeta_{s_T,n}}
\prod_{t=T-1}^0
	\frac{
	1-\beta
	}
	{(1-\beta)+\beta\zeta_{s_t,n}}
=
	\frac{
	\beta\zeta_{s_T,t}
	}
	{(1-\beta)}
\prod_{t=T}^0
	\frac{
	1-\beta
	}
	{(1-\beta)+\beta\zeta_{s_t,n}}.
\label{eq:Cinternal}
\ee
By Lemma~\ref{lem:gamma}, 
the coefficients of the form~(\ref{eq:Cleaf}) tend to one
while 
the coefficients of the form~(\ref{eq:Cinternal}) tend to zero,
therefore, 
a.s., as $n\to\infty$,
$$P^*_D(j|x_{-D+1}^n)-P_{e,s_D,n+1|n}
=P^*_D(j|x_{-D+1}^n)-
\frac{a_{s_D,n+1}(j)-1/2}{m/2+M_{s_D,n+1}-1}
\to 0.$$
Moreover, the ergodic theorem 
for Markov chains~\cite{chung:book,meyn-tweedie:book2} implies
that for any context $s$ of length $D$,
a.s., as $n\to\infty$,
\be
\frac{a_{s,n}(j)}{M_{s,n}}
=\frac{a_{s,n}(j)}{n}
\cdot
\frac{n}{M_{s,n}}
\to 
\frac{\pi(sj)}{\pi(s)}=
\theta_s(j),
\label{eq:ergodic}
\ee
where with a slight abuse of notation we write
$\pi(s)$ for the probability assigned by the
stationary distribution of the chain to
the string corresponding to a context $s$.
The proof is completed upon noting that,
since there are only finitely many contexts 
$s$ of length no more than $D$, the convergence
in~(\ref{eq:ergodic})
occurs uniformly in $s$.
\qed

\medskip

\noindent
{\em Proof of Theorem~\ref{thm:modelC}. }
Consider an alternative model $T\in\clT(D)$, 
different from the true minimal model $T^*$.
We will show that the posterior probability
of $T$ will be strictly smaller than that
of $T^*$, eventually almost surely (a.s.), as the
size $n$ of the sample 
increases. Since
there are only finitely many models in 
$\clT(D)$, this suffices to prove the theorem.

We consider two cases, which are not
necessarily mutually exclusive: Since $T\neq T^*$,
either there is a leaf $t\in T$ such that the 
collection $T^*(t)$ of contexts in $T^*$ that are 
descendants of $t$ in $T$ is nonempty; or there
is a $t\in T^*$ such that the
collection $T(t)$ of contexts in $T$ 
that are descendants of $t$ in $T^*$
is nonempty.

\medskip

\noindent
{\em Case 1. } Let $t\in T$ be at level $d<D$,
and such that $T^*(t)\neq\Lambda$.
We will show that the posterior of the union
$T\cup T^*(t)$, which consists of $T$ together
with the subtree $T^*(t)$ starting at~$t$,
satisfies,
$\pi(T\cup T^*(t)|x)>\pi(T|x)$,
that is, 
$P(x|T\cup T^*(t))\pi_D(T\cup T^*(t))
>P(x|T)\pi_D(T)$, or equivalently, using Lemma~\ref{lem:integrate},
$$\sum_{s\in T\cup T^*(t)}\log P_e(a_s) + \log\pi_D(T\cup T^*(t)) 
>\sum_{s\in T}\log P_e(a_s) + \log\pi_D(T),
\qquad\mbox{eventually, a.s.},
$$
which, using Lemma~\ref{lem:unions} 
is,
$$
\sum_{s\in T^*(t)}\log P_e(a_s) 
-\log P_e(a_t)
+\log\pi_{D-d}(T^*(t)) -\log\beta
> 0,
\qquad\mbox{eventually, a.s.,}
$$
or, equivalently,
\begin{align*}
\sum_{s\in T^*(t)}\log\left(\frac{P_e(a_s)}{\prod_j\theta_s(j)^{a_s(j)}}\right)
-\log & \left(\frac{P_e(a_t)}{\prod_j(a_t(j)/M_t)^{a_t(j)}}\right)
-\log\left(\frac{\prod_j(a_t(j)/M_t)^{a_t(j)}}{\prod_{s\in T^*(t)}\prod_j
\theta_s(j)^{a_s(j)} }\right)\\
&
+\log\pi_{D-d}(T^*(t)) -\log\beta
> 0,
\qquad\mbox{eventually, a.s.}
\end{align*}
Using~(\ref{eq:PeLB})
of Lemma~\ref{lem:KT} as in the
proof of Theorem~\ref{thm:redundancy},
the first term above is bounded
below by,
$$
-\sum_{s\in T^*(t)}\left(
\frac{m-1}{2}\log M_s-\log m\right).
$$
Similarly, using~(\ref{eq:PeUB})
of Lemma~\ref{lem:KT},
the second term above is bounded
below by,
$$
\frac{m-1}{2}\log\frac{M_t}{2\pi}+\log\frac{\pi^{m/2}}{\Gamma(m/2)}.
$$
And writing
$\hat{p}_t(j)=a_t(j)/M_t$
as in the proof of Proposition~\ref{prop:MLE},
the third term above equals,
$$
M_tH(\hat{p}_t)
+\sum_{s\in T^*(t)}\sum_j
a_s(j)\log \theta_s(j).
$$
In fact, we will prove the stronger
fact that
$\liminf_n(1/n)[\log \pi(T\cup T^*(t)|x)-\log \pi(T|x)]>0$,
a.s., for which, combining the above
expressions, it suffices to show that,
a.s.,
$$
\liminf_{n\to\infty}
\left\{
\frac{m-1}{2n}
\log M_t
-
\frac{m-1}{2n}
\sum_{s\in T^*(t)} \log M_s
+\frac{M_t}{n}H(\hat{p}_t)
+\frac{1}{n}\sum_{s\in T^*(t)}\sum_j
a_s(j)\log \theta_s(j)\right\}
> 0.
$$
Since $M_t$ is always no greater than $n$, 
the first term
goes to zero as $n\to\infty$,
and using Jensen's inequality as
in the proof of Theorem~\ref{thm:redundancy}
we also have that the second term is bounded
below by
$[|T^*(t)|(m-1)/(2n)]\log(n/|T^*(t)|)=O((\log n)/n)$,
which also goes to zero a.s. as $n\to\infty$. 
Therefore, it suffices to show that,
$$
\liminf_{n\to\infty}
\left\{
\frac{M_t}{n}H(\hat{p}_t)
+
\sum_{s\in T^*(t)}
\frac{M_s}{n}
\sum_j
\frac{a_s(j)}{M_s}\log \theta_s(j)\right\}
> 0,\qquad\mbox{a.s.}
$$
Using the same notation as 
in equation~(\ref{eq:ergodic}), 
in the proof of Theorem~\ref{thm:pred}, 
the ergodic theorem~\cite{chung:book,meyn-tweedie:book2}
implies that 
$M_s/n\to\pi(s)$ and
$a_s(j)/M_s\to\theta_s(j)$,
a.s., as $n\to\infty$,
so that the 
above $\liminf$ is actually
a limit, which equals,
\be
\pi(t)
H(\theta_t)
-
\sum_{s\in T^*(t)}
\pi(s)
H(\theta_s).
\label{eq:Dentropies}
\ee
And the strict concavity
of the entropy implies that,
$$
\sum_{s\in T^*(t)}
\frac{\pi(s)}{\pi(t)}
H(\theta_s)
\leq
H\left(\sum_{s\in T^*(t)}
\frac{\pi(s)}{\pi(t)}\;
\theta_s\right)=
H(\theta_t),
$$
with equality only 
if all $\theta_s$ are equal,
which is ruled out by the
assumption that the model
$T^*$ is minimal.
This implies that the difference
in~(\ref{eq:Dentropies}) is 
strictly positive and
completes this case. 

\medskip

\noindent
{\em Case 2. } 
Let $t\in T^*$ be a leaf at level
$d\geq 1$ such that the
collection $T(t)$ of contexts in $T$ 
that are descendants of $t\in T^*$
is nonempty, so that $T$ can be
expressed as the union $T^p\cup T(t)$
of the tree $T^p$ which is $T$
pruned at $t$, and $T(t)$.
Since $\pi(t)>0$ by assumption,
we must have that $\pi(s)>0$ for 
at least one $s\in T(t)$.
If this holds for exactly only
one $s\in T(t)$, then a 
simple calculation shows that
$\pi(T|x)=\pi(T^p|x)$. Therefore,
we assume that 
$\pi(s)>0$ for at least two
contexts $s\in T(t)$, and 
we will show that 
$\pi(T^p|x)>\pi(T|x)=\pi(T^p\cup T(t)|x)$,
i.e, that
$P(x|T^p)\pi_D(T^p)
>P(x|T^p\cup T(t))\pi_D(T(t))$. 
As in Case~1, 
using Lemmas~\ref{lem:integrate}
and~\ref{lem:unions},
this is easily seen to be
the same as,
$$
-\sum_{s\in T(t)}\log P_e(a_s) 
+\log P_e(a_t)
-\log\pi_{D-d}(T(t)) +\log\beta
> 0,
\qquad\mbox{eventually, a.s.,}
$$
or, equivalently,
\begin{align*}
\sum_{s\in T(t)}\log\left(\frac
	{\prod_j(a_s(j)/M_s)^{a_s(j)}}
	{P_e(a_s)}
	\right)
+\log & \left(\frac{P_e(a_t)}{\prod_j\theta_t(j)^{a_t(j)}}\right)
-\log\left(\frac
{\prod_{s\in T(t)}\prod_j (a_s(j)/M_s)^{a_s(j)} }
{\prod_j\theta_t(j)^{a_t(j)}}
\right)\\
&
-\log\pi_{D-d}(T(t)) +\log\beta
> 0,
\qquad\mbox{eventually, a.s.}
\end{align*}
Note that in the sums and products
over $s\in T(t)$, we can (and do)
restrict attention to only those
$s$ with $\pi(s)>0$.
Again, using~(\ref{eq:PeLB})
of Lemma~\ref{lem:KT} like in the
proof of Theorem~\ref{thm:redundancy},
the second term above is bounded
below by,
$$
-\frac{m-1}{2}\log M_t-\log m,
$$
similarly, using~(\ref{eq:PeUB})
of Lemma~\ref{lem:KT},
the first term is bounded
below by,
$$
\sum_{s\in T(t)}
\left(
\frac{m-1}{2}\log\frac{M_s}{2\pi}+
\log\frac{\pi^{m/2}}{\Gamma(m/2)}
\right),
$$
and noting that
$a_t(j)=\sum_{s\in T(t)}a_s(j)$ for all $j$,
the third term is actually equal to,
$$
-\sum_{s\in T(t)}M_s
	\sum_j \frac{a_s(j)}{M_s}\log\frac{a_s(j)}{M_s}
+\sum_j a_t(j)\log\theta_t(j)
=
-\sum_{s\in T(t)}M_s D(\hat{p}_s\|\theta_t),
$$
where, as before, 
$\hat{p}_s(j)=a_s(j)/M_s$.
Combining the above
expressions, it suffices to show that,
eventually a.s.,
\begin{align*}
\sum_{s\in T(t)}
M_s
D(\hat{p}_s\|\theta_t)
<
\frac{m-1}{2}\left(\sum_{s\in T(t)} \log M_s-\log M_t\right)
&+|T(t)|\log\frac{\pi^{1/2}}{
2^{(m-1)/2}
\Gamma(m/2)}\\
&-\log\pi_{D-d}(T(t))
+\log\beta,
\end{align*}
and since the last three terms above
are uniformly bounded in $n$, it
suffices to show that,
\begin{align}
\limsup_{n\to\infty}
\left\{
\frac{1}{\log n}
\sum_{s\in T(t)}
M_s
D(\hat{p}_s\|\theta_t)
-\frac{m-1}{2\log n}\left(\sum_{s\in T(t)} \log M_s-\log M_t\right)
\right\}
<0,
\qquad\mbox{a.s.}
\label{eq:LILtarget}
\end{align}
For the first term above we have,
\ben
	\sum_{s\in T(t)}
	M_s
	D(\hat{p}_s\|\theta_t)
&\leqa& 
	\sum_{s\in T(t)}
	M_s
	\sum_j\frac{(\hat{p}_s(j)-\theta_t(j))^2}
	{\theta_t(j)}\\
&=& 
	\sum_{s\in T(t)}
	\frac{1}{M_s}
	\sum_j\frac{(a_s(j)-\theta_t(j)M_s)^2}
	{\theta_t(j)}\\
&\eqb& 
	\sum_{s\in T(t)}
	\frac{1}{M_s}
	\sum_j\frac{[(a_s(j)-n\pi(sj))-\theta_t(j)(M_s-n\pi(s))]^2}
	{\theta_t(j)}\\
&=& 
	(\log\log n)
	\sum_{s\in T(t)}
	\frac{n}{M_s}
	\sum_j
	\frac{1}
	{\theta_t(j)}
	\left[
	\frac{a_s(j)-n\pi(sj)}{\sqrt{n\log\log n}}
	-\theta_t(j)
	\frac{M_s-n\pi(s)}{\sqrt{n\log\log n}}\right]^2,
\een
where
$(a)$ follows from the well-known bound 
for the relative entropy
in terms of the $\chi^2$ distance~\cite{gibbs:02},
and $(b)$ follows from the fact that, since $T^*$
is minimal and $s$ is a descendant of $t\in T^*$,
we have $\pi(sj)/\pi(s)=\theta_t(j)$.
By the law of the iterated
logarithm~\cite{chung:book}
applied to the chain 
$\{Z_n=X_{n-D+1}^n;n\geq 1\}$,
each of the two fractions in the
above square brackets is $O(1)$
a.s., and by the ergodic theorem 
so is $n/M_s$, so that the entire
expression, as $n\to\infty$,
\be 
\sum_{s\in T(t)}
	M_s
	D(\hat{p}_s\|\theta_t)
=O(\log\log n),\qquad\mbox{a.s.}
\label{eq:LILtarget2}
\ee
Moreover, the second term in~(\ref{eq:LILtarget})
equals,
\be
\frac{m-1}{2\log n}\left(\sum_{s\in T(t)} \log \frac{M_s}{n}
-\log \frac{M_t}{n}
+(|T^+|-1)\log n\right),
\label{eq:logn1}
\ee
where $|T^+|\geq 2$ denotes
the number of $s\in T(t)$ 
such that $\pi(s)>0$.
Since, by the ergodic theorem,
$M_s/n\to\pi(s)$ and $M_t/n\to\pi(t)$,
a.s., as $n\to\infty$, 
the above expression converges a.s.\
to,
\be
\frac{(m-1)(|T^+|-1)}{2}>0.
\label{eq:logn2}
\ee
Combining this with~(\ref{eq:LILtarget2})
shows that~(\ref{eq:LILtarget}) holds,
completing the proof of this case
and proving the theorem.
\qed

In order to prove Theorem~\ref{thm:modelC2}
we need the simple asymptotic result of Proposition~\ref{prop:SMBT},
which can be seen as a version 
of the Shannon-McMillan-Breiman theorem 
in this setting; cf.~\cite{cover:book,barron:1}.
For an ergodic chain $\{X_n\}$ with minimal model
$T^*$, parameters $\theta^*$ and stationary distribution
$\pi$ we will use the following notation. As before,
for any context $s$ in $T^*$ (or in any other model),
with a slight abuse of notation we write $\pi(s)$
for the stationary probability of the string
corresponding to $s$. 
Similarly, we write $\theta^*_s(j)$ for the conditional
probability that $j$ will follow context $s$; and
if $s$ is not in $T^*$, we simply take 
$\theta^*_s(j)$ to be the ratio $\pi(sj)/\pi(s)$.

\begin{proposition}
\label{prop:SMBT}
Suppose $\{X_n\}$ is an ergodic
chain with minimal model $T^*\in\clT(D)$,
associated parameters $\theta^*=\{\theta_s^*;s\in T^*\}$,
and stationary distribution $\pi$.
Then the marginal likelihood $P(X_1^n|X_{-D+1}^0,T)$ 
with respect to any model $T\in\clT(D)$
decays exponentially with
the sample size~$n$: 
As $n\to\infty$, 
$$-\frac{1}{n} P(X_1^n|X_{-D+1}^0,T)\to \bar{H}(X|T)
\geq 0,\qquad\mbox{a.s.},$$
where $\bar{H}(X|T)$ denotes the entropy rate
functional,
$$
\bar{H}(X|T)=-\sum_{s\in T}
\pi(s)\sum_{j\in A} \theta_s^*(j)\log\theta_s^*(j).$$
Moreover, $\bar{H}(X|T^*) < \bar{H}(X|T)$ for any 
proper subtree $T$ of $T^*$. 
\end{proposition}

\noindent
{\em Proof. }
This will be seen to be a simple consequence of the bounds in 
Lemma~\ref{lem:KT}. From Lemma~\ref{lem:integrate}
we know that the marginal likelihood, 
$$
\log P(X_1^n|X_{-D+1}^0,T)
=
\sum_{s\in T}\log P_s(a_s),$$
and from the bounds in  
Lemma~\ref{lem:KT} this can be expressed,
$$
\log P(X_1^n|X_{-D+1}^0,T)
=
\sum_{s\in T:M_s\neq 0}\left[
\sum_{j\in A}a_s(j)\log\frac{a_s(j)}{M_s}
-\frac{m-1}{2}\log M_s+A_n\right],
$$
where the (possibly random) constants $A_n$
only depend on $m$ and they are $O(1)$ a.s.
Therefore, noting also that $0\leq M_s\leq n$ a.s.,
we have,
$$
-\frac{1}{n}\log P(X_1^n|X_{-D+1}^0,T)
=
-\sum_{s\in T}\frac{M_s}{n}\left[
\sum_{j\in A}\frac{a_s(j)}{M_s}\log\frac{a_s(j)}{M_s}\right]
+O\Big(\frac{\log n}{n}\Big),
$$
and recalling, as noted in equation~(\ref{eq:ergodic})
in the proof of Theorem~\ref{thm:pred}
above, that $M_s/n\to\pi(s)$ and $a_s(j)/M_s\to\theta^*_s(j)$, a.s.,
for all $s\in T$ and $j\in A$, the asymptotic result follows. 

The nonnegativity of $\bar{H}(X|T)$ is obvious from its
definition. Finally, suppose that $T$ is a proper subtree
of $T^*$, 
and for each $t\in T$
let $T^*(t)$ denote the collection of descendants of $t$
that are leaves of $T^*$ (or $T^*(t)=\{t\}$ if $t$ is a
leaf of both $T$ and $T^*$).
We can write,
\ben
\bar{H}(X|T^*)
&=&
-\sum_{t\in T}\sum_{s\in T^*(t)}
\pi(s)\sum_{j\in A} \theta_s^*(j)\log\theta_s^*(j)\\
&=&
-\sum_{t\in T}
\pi(t)
\sum_{j\in A} 
\sum_{s\in T^*(t)}
\frac{\pi(s)}{\pi(t)}
	\theta_s^*(j)\log\left(\frac{\frac{\pi(s)}{\pi(t)}\theta_s^*(j)}
	{\frac{\pi(s)}{\pi(t)}}\right),
\een
and using the 
log-sum inequality~\cite{cover:book},
\ben
\bar{H}(X|T^*)
\leq
-\sum_{t\in T}
\pi(t)
\sum_{j\in A} 
\left(
\sum_{s\in T^*(t)}
\frac{\pi(s)}{\pi(t)}
	\theta_s^*(j)
	\right)\log\left(
	\frac{\sum_{s\in T^*(t)} 
	\frac{\pi(s)}{\pi(t)}\theta_s^*(j)}
	{\sum_{s\in T^*(t)}
	\frac{\pi(s)}{\pi(t)}}\right),
\een
Now, from the definitions we obviously have
$\sum_{s\in T^*(t)}\pi(s)=\pi(t)$ and
$\sum_{s\in T^*(t)}\pi(s)\theta^*_s(j)=\pi(t)\theta^*_t(j)$,
so that,
\ben
\bar{H}(X|T^*)
\leq
-\sum_{t\in T}
\pi(t)
\sum_{j\in A} 
	\theta_t^*(j)\log\theta_t^*(j)=
\bar{H}(X|T).
\een
Finally, we would have equality in~$(a)$ only if 
for all $j$, the
$\theta_s(j)$ were independent of $s$, but
that contradicts the minimality of $T^*$, so
the inequality is necessarily strict.
\qed

\noindent
{\em Proof of Theorem~\ref{thm:modelC2}. }
We will show that $\pi(T|X_{-D+1}^n)\to 0$ a.s., 
for all $T\in\clT(D)$, $T\neq T^*$. 
We consider two cases.

\medskip

\noindent
{\em Case 1. } First, suppose that there
is an internal node $t\in T^*$ which is 
a leaf of $T$. By successive operations of 
adding and/or removing subtrees from $T$ as in
the two cases considered in the proof of 
Theorem~\ref{thm:modelC}, we see that 
the tree $T'$ that is exactly the same
as $T^*$ but pruned at $t$, has posterior
probability greater than $T$ eventually a.s.
So we can assume, without loss of generality,
that $T$ is of that form. 

Then, the posterior
probability of $T$ can easily be bounded above,
\ben
\pi(T|X_{-D+1}^n)
&=&
	\frac{P(X_1^n|X_{-D+1}^0,T)\pi(T)}
	{P_D^*(x)}\\
&=&
	\frac{P(X_1^n|X_{-D+1}^0,T)\pi(T)}
	{\sum_{T'} P(X_1^n|X_{-D+1}^0,T')\pi(T')}\\
&\leq&
	\frac{\pi(T)}{\pi(T^*)}
	\cdot
	\frac{P(X_1^n|X_{-D+1}^0,T)}
	{P(X_1^n|X_{-D+1}^0,T^*)},
\een
so that, using Proposition~\ref{prop:SMBT},
we have a.s.\ as $n\to\infty$,
\ben
\frac{1}{n}\log\pi(T|X_{-D+1}^n)
&=&
	\frac{1}{n}\log\left(\frac{\pi(T)}{\pi(T^*)}\right)
	-\frac{1}{n}\log
	P(X_1^n|X_{-D+1}^0,T^*)
	+\frac{1}{n}\log P(X_1^n|X_{-D+1}^0,T)\\
&\to&
	\bar{H}(X|T^*)-\bar{H}(X|T),
\een
which is strictly negative because of our assumption
that $T$ is a proper subtree of $T^*$. Therefore,
$\pi(T|X_{-D+1}^n)\to 0$ a.s., exponentially fast
as $n\to\infty$.

\medskip

\noindent
{\em Case 2. } 
Alternatively, if no internal 
node $t$ of $T^*$ is a leaf of $T$,
then $T^*$ is a proper subtree of $T$. 
Again by successively repeating the 
pruning operation as in Case~2 in the
proof of Theorem~\ref{thm:modelC},
which increases the posterior probability
of $T$ eventually a.s., 
we can assume without loss of generality
that $T$ consists of exactly $T^*$
together with $m$ additional leaves
$\{tk\;;\;k\in A\}$ stemming from
a specific $t\in T^*$.
Then proceeding as in Case~1 above
we have,
\ben
\log \pi(T|X_{-D+1}^n)
\leq
	\log\left(\frac{\pi(T)}{\pi(T^*)}\right)
	+\log P(X_1^n|X_{-D+1}^0,T)
	-\log P(X_1^n|X_{-D+1}^0,T^*),
\een
and using Lemma~\ref{lem:integrate} and
the bounds from Lemma~\ref{lem:KT},
\ben
\log \pi(T|X_{-D+1}^n)
&\leq&
	\sum_{s\in T:M_s\neq 0}
	\left[\sum_{j=0}^{m-1}a_s(j)\log\frac{a_s(j)}{M_s}
	-\frac{m-1}{2}\log M_s
	-C_1
	\right]\\
&&-
	\sum_{s\in T^*:M_s\neq 0}
	\left[
	\sum_{j=0}^{m-1}a_s(j)\log\frac{a_s(j)}{M_s} 
	-\frac{m-1}{2}\log M_s-C_2\right] +C_3,
\een
where
$C_1=\log\left(\frac{\sqrt{2\pi}}{2^{m/2}\Gamma(m/2)}\right)$,
$C_2=\log m$ and $C_3= \log\left(\frac{\pi(T)}{\pi(T^*)}\right)$.
Since $t\in T^*$ and we assume that $\pi(t)>0$, by the ergodic
theorem we know that $M_t\neq 0$ eventually a.s.,
therefore, we have,
\ben
\log \pi(T|X_{-D+1}^n)
&\leq&
	\sum_{k\in A:M_{tk}\neq 0}
	\left[\sum_{j=0}^{m-1}a_{tk}(j)\log\frac{a_{tk}(j)}{M_{tk}}
	-\frac{m-1}{2}\log M_{tk}
	\right]\\
&&
	-
	\left[\sum_{j=0}^{m-1}a_t(j)\log\frac{a_t(j)}{M_t} 
	-\frac{m-1}{2}\log M_t\right] +C_4\\
&\leq&
	\sum_{k\in A:M_{tk}\neq 0}
	\left[M_{tk}\sum_{j=0}^{m-1}\frac{a_{tk}(j)}{M_{tk}}
	\log\frac{a_{tk}(j)}{M_{tk}}
	\right]
	-\sum_{j=0}^{m-1}a_t(j)\log\theta^*_t(j)
	\\
&&
	-\frac{m-1}{2}\left(
	\sum_{k\in A:M_{tk}\neq 0}
	\log M_{tk}
	-\log M_t\right)+C_4,
\een
eventually a.s., for a finite constant $C_4$,
and where we used, as in the proofs of Theorems~\ref{thm:redundancy}
and~\ref{thm:modelC} the fact that the empirical frequencies
$\hat{p}_t(j)=a_t(j)/M_t$ maximise the likelihood. Now recalling 
that $\sum_k a_{tk}(j)=a_t(j)$, we have, eventually a.s.,
$$
\log \pi(T|X_{-D+1}^n)\leq
\sum_{k\in A:M_{tk}\neq 0} M_{tk} D(\hat{p}_{tk}\|\theta^*_t)
	-\frac{m-1}{2}\left(
	\sum_{k\in A:M_{tk}\neq 0}
	\log M_{tk}
	-\log M_t\right)+C_4,
$$
and by exactly the same argument as the one that led 
to~(\ref{eq:LILtarget2}) in the proof of Theorem~\ref{thm:modelC},
$$
\log \pi(T|X_{-D+1}^n)\leq
	-\frac{m-1}{2}\left(
	\sum_{k\in A:M_{tk}\neq 0}
	\log M_{tk}
	-\log M_t\right)
	+O(\log\log n),\qquad\mbox{a.s.}
$$
Finally, again by the same argument that led to~(\ref{eq:logn1})
and~(\ref{eq:logn2}) in the proof of Theorem~\ref{thm:modelC},
we have, that,
$$\log \pi(T|X_{-D+1}^n)\leq
	-C_5\log n
	+o(\log n),\qquad\mbox{a.s.},
$$
which implies that 
$\log \pi(T|X_{-D+1}^n)\to -\infty$ and hence
$\pi(T|X_{-D+1}^n)\to 0$ a.s., as $n\to\infty$,
completing the proof.
\qed

Although superficially somewhat technical, 
Theorems~\ref{thm:consistency} and~\ref{thm:normality} 
proved next are simple consequences of the exact 
form~(\ref{eq:full-cond}) of the full conditional 
density of $\theta$ given $x_{-D+1}^n$ and $T$, 
combined with Theorem~\ref{thm:modelC2} and with 
some simple convergence properties of the Dirichlet 
distribution~\cite{geyer:13}.

\medskip

\noindent
{\em Proof of Theorem~\ref{thm:consistency}. }
Let $x=x_{-D+1}^\infty$ be a semi-infinite
sample realisation. For each $n$, the
posterior distribution $\pi(\theta,T|x_{-D+1}^n)$
can formally be described as probability measure 
$\mu$
on the space $\clS$ consisting of elements
$(\theta,T)$, where $T\in\clT(D)$
and $\theta=\{\theta_s;s\in T\}$
with each $\theta_s\in[0,1]^m$.
We endow $\clS$ with the $\sigma$-algebra
$\clJ$ consisting of all sets $S$ of the
form,
$$
S=\bigcup_{T\in\clT(D)}\big(B_T\times\{T\}\big),
\qquad B_T\in\clB^{m|T|},
$$
where each $\clB^{m|T|}$ denotes the Borel $\sigma$-algebra
of $[0,1]^{m|T|}$. Then the probability of
any such $S$ can be decomposed as,
$$
\mu(S)=
\sum_{T\in\clT(D)}\pi(T|x_{-D+1}^n)
\pi(B_T|T,x_{-D+1}^n),
$$
and from Theorem~\ref{thm:modelC2} 
we know that, 
for almost all realisations $x$,
$\pi(T|x_{-D+1}^n)$ asymptotically 
concentrates on $T^*$,
so that,
$$\lim_n\mu(S)=\lim_n
\pi(B_{T^*}|T^*,x_{-D+1}^n).
$$
Therefore,
writing $\theta^{(n)}$ for a random
vector with distribution
$\pi(\cdot|x_{-D+1}^n,T^*)$,
in order to establish
the required result it suffices
to show that, for any $B\in\clB^{m|T^*|}$,
\ben
\lim_n\BBP\left(\theta^{(n)}\in B\Big|x_{-D+1}^n,T^*\right)
=\IND\{\theta^*\in B\},\qquad\mbox{for almost all}\;x,
\een
or, equivalently, 
that $\theta^{(n)}$ converges
in probability to $\theta^*$, for almost all $x$,
where $\IND\{\cdot\}$ denotes the indicator
function of the event $\{\cdot\}$.

Given the sample string $x_{-D+1}^n$ 
up to time $n$, as in the proof 
of Theorem~\ref{thm:pred} we write 
$a_{s,n}$ and $M_{s,n}$ for the induced
count vectors and 
$\hat{p}_{s,n}(j)=a_{s,n}(j)/M_{s,n}$, 
$j\in A$ for the corresponding
empirical frequencies corresponding
to each context $s\in T^*$.
By the same reasoning as
in equation~(\ref{eq:ergodic})
earlier, we have that
$\hat{p}_{s,n}\to \theta^*_s(j)$ and
$M_{s,n}/n\to\pi(s)$ a.s., as $n\to\infty$,
for each $s\in T^*$.
Let $\clA$ denote the set of all realisations $x$
such that the result of 
Theorem~\ref{thm:modelC2} as well as 
all the above asymptotics hold,
so that $\clA$ has probability~1.

Choose and fix any one of the (almost all)
realisations $x=x_{-D+1}^\infty\in\clA$ 
for the remainder of the proof.
As noted in equation~(\ref{eq:full-cond}),
the distribution $\pi(\cdot|x_{-D+1}^n,T^*)$ 
of $\theta^{(n)}$ has
a density $f_n(\theta)$ with respect to Lebesgue
measure, given by the product,
\be
f_n(\theta)=\prod_{s\in T^*}f_{n,s}(\theta_s),
\label{eq:productD}
\ee
where each $f_{n,s}$ denotes
the Dir$(a_s(0)+1/2,\ldots,a_s(m-1)+1/2)$
density, so that, in particular, it is easy to
compute the corresponding means,
\be
\bar{\theta}^{(n)}_s(j)
:=\BBE\left(\theta^{(n)}_s(j)\Big |x_{-D+1}^n,T^*\right)
=\frac{a_{s,n}(j)+1/2}{M_s+m/2}
\to\theta^*_s(j),\qquad\mbox{as }n\to\infty,
\label{eq:centering2}
\ee
and variances,
$$
\VAR\left(\theta^{(n)}_s(j)) \Big |x_{-D+1}^n,T^*\right)
=\frac
{(a_{s,n}(j)+1/2)(M_s-a_{s,n}(j)+(m-1)/2)}
{(M_s+m/2)^2(M_s+m/2+1)}
\to0,\qquad\mbox{as }n\to\infty.
$$
Then a simple application of Chebyshev's inequality
implies that $\theta^{(n)}$ converges
in probability to $\theta^*$, completing the proof.
\qed

\noindent
{\em Proof of Theorem~\ref{thm:normality}. }
We follow the same reasoning and adopt the same
notation as in the first part of the proof 
of Theorem~\ref{thm:consistency},
and again we choose and fix an arbitrary
$x=x_{-D+1}^\infty\in\clA$.
Then, for each $n$, the density $f_n(\theta)$
of $\pi(\theta|x_{-D+1^n},T^*)$ is given by
by the product in~(\ref{eq:productD}),
and the claim~(\ref{eq:centering}) has
already been established in~(\ref{eq:centering2}).

In order to establish the asymptotic
normality of $\theta^{(n)}$,
for each $s\in T^*$, let
$\phi_{J_s}(\cdot)$ denote the $N(0,J_s)$ density
on $\RL^m$, with $J_s$ defined in~(\ref{eq:Js}).
Since the collection of all sets of the form,
$$
\prod_{s\in T^*}
\Big(
[0,x_s(0))\times[0,x_s(1))\times\cdots\times[0,x_s(m-1))
\Big),
$$
for $x_s(j)\in[0,1]$, $s\in T^*$, $j\in A$, form
a $\pi$-system for the Borel $\sigma$-algebra
of $[0,1]^{m|T^*|}$, and also since 
for all $n$ the components $\theta^{(n)}_s$ of $\theta^{(n)}$
for different $s\in T^*$ are independent,
in order to prove the theorem it suffices
to show~\cite{billingsley:cpm} that for each $s\in T^*$,
\be
\frac{1}{\sqrt{n}}f_{s,n}
\left(\frac{z}{\sqrt{n}}+\bar{\theta}^{(n)}_s\right)
\to\phi_{J_s}(z),
\qquad\mbox{as}\;n\to\infty,
\label{eq:targetD}
\ee
where the convergence is uniform on compact subsets
of $\RL^m$.

From Theorems~4.2 and~4.3 of~\cite{geyer:13}
we have that, uniformly on compact sets,
\be
\frac{1}{\sqrt{\nu_{s,n}}}f_{s,n}
\left(\frac{z}{\sqrt{\nu_{s,n}}}+\tilde{\theta}^{(n)}_s\right)
\to\phi_{I_s}(z),
\qquad\mbox{as}\;n\to\infty,
\label{eq:geyer}
\ee
where $\nu_{s,n}=M_{s,n}+m/2$, 
$$\tilde{\theta}^{(n)}_s(j)=\frac{a_{s,n}(j)-1/2}{M_s-m/2},
\qquad j\in A,$$
and
$I_s=\Theta^*_s-(\theta^*_s)^t(\theta^*_s).$
But from our assumptions we have that $I_s=\pi(s)J_s$
and that, as $n\to\infty$,
$\nu_{s,n}/n\to\pi(s)$ and
$\tilde{\theta}^{(n)}_s-
\bar{\theta}^{(n)}_s\to 0$.
These together with~(\ref{eq:geyer})
and the continuous mapping theorem~\cite{billingsley:cpm}
imply~(\ref{eq:targetD}) as required.
\qed

\noindent
{\em Proof of Theorem~\ref{thm:shtarkov}. }
The proof follows roughly along the same lines
as the one for the special case of binary
data and $\beta=1/2$ given 
in~\cite{willems-shtarkov-tjalkens:pre,ctw-4:93},
which in turn is a generalisation of Shtarkov's
original argument in~\cite{shtarkov:87}.

For each $n$, each string $x_1^n$, each initial
context $x_{-D+1}^n$, and any $T\in\clT(D)$,
we denote by $\Sigma(x_{-D+1}^0,x_1^n,T)$
the expression,
\ben
\Sigma(x_{-D+1}^0,x_1^n,T)
&=&
\sum_{s\in T:M_s\neq 0}
\left(
\frac{m-1}{2}
\log \frac{M_s}{2\pi}
+
\log\Big(\frac{\pi^{m/2}}{\Gamma(m/2)}\Big)
\right)\\
&=&
	\sum_{s\in T:M_s\neq 0}\left(
	\frac{m-1}{2}\log M_s
	+\log\Big(\frac{\sqrt{2\pi}}{
	2^{m/2}
	\Gamma(m/2)}\Big)\right),
\een
where $M_s$ are the sums of the count vectors $a_s$ 
corresponding to $x_{-D+1}^n$,
and we define a 
(conditional) probability measure $\mu$
on $A^n$ as,
$$\mu(x_1^n|x_{-D+1}^0)=\frac{1}{Z(x_{-D+1}^0)}\times
\max_{T\in\clT(D)}\sup_{\phi\in\Omega(T,m)}
\left[
\frac{P(x_1^n|x_{-D+1}^0,\theta(\phi),T)}
{\exp\left\{
\Sigma(x_{-D+1}^0,x_1^n,T)
-\log \pi_D(T;\beta)
\right\}}
\right],
$$
where 
$Z(x_{-D+1}^0)$ is simply the normalising constant,
\begin{align*}
&Z(x_{-D+1}^0)\\
&=\sum_{y_1^n\in A^n}
	\max_{T\in\clT(D)}
	\left[
	\frac{
	\sup_{\phi\in\Omega(T,m)}
	P(y_1^n|x_{-D+1}^0,\theta(\phi),T)}
	{\exp\left\{
	\Sigma(x_{-D+1}^0,y_1^n,T)
	-\log\pi_D(T;\beta)
	\right\}}
	\right].
\end{align*}
As we saw in Proposition~\ref{prop:MLE},
the supremum in the numerator above is achieved
by the choice of parameters 
$\hat{\theta}_s=a_s/M_s$, for all $s\in T$, so that,
$$
	\sup_{\phi\in\Omega(T,m)}
	P(y_1^n|x_{-D+1}^0,\theta(\phi),T)
=
	P(y_1^n|x_{-D+1}^0,\hat{\theta},T)
=
	\prod_{s\in T}
	\prod_{j\in A}\Big(\frac{a_s(j)}{M_s}\Big)^{a_s(j)},
$$
hence,
\begin{align*}
Z(&x_{-D+1}^0)=
\sum_{y_1^n\in A^n}
	\max_{T\in\clT(D)}\\
&
	\exp
	\left\{
	\sum_{s\in T:M_s\neq 0}
	\left[
	\sum_{j\in A}
	a_s(j)\log\Big(\frac{a_s(j)}{M_s}\Big)
	-\frac{m-1}{2}\log \frac{M_s}{2\pi}
	-\log\Big(\frac{\pi^{m/2}}{\Gamma(m/2)}\Big)
	\right]
	+
	\log \pi_D(T;\beta)
	\right\}.
\end{align*}
Further, using the bound~(\ref{eq:PeUB}) in Lemma~\ref{lem:KT},
and the expression for the marginal likelihood 
in Lemma~\ref{lem:integrate}, we have,
\begin{align}
Z(x_{-D+1}^0)
&\geq
\sum_{y_1^n\in A^n}
	\max_{T\in\clT(D)}
	\exp\left\{
	\sum_{s\in T:M_s\neq 0}
	\log P_e(a_s)
	+\log \pi_D(T;\beta)
	\right\}\nonumber\\
&=
\sum_{y_1^n\in A^n}
	\max_{T\in\clT(D)}
	\left[P(y_1^n|x_{-D+1}^0,T)
	\pi_D(T;\beta)\right].
\label{eq:partition}
\end{align}

Now, by the definition of $\mu$, and noting that a likelihood
ratio cannot be uniformly smaller than~1, after some simple
algebra we have,
\begin{align*}
-\min_{T\in\clT(D)}
&
	\inf_{\phi\in\Omega(T,m)}\min_{x_1^n\in S_n}\\
&
	\hspace{0.6in}
	\Bigg\{
	\log Q_n(x_1^n)
	-\log P(x_1^n|x_{-D+1}^0,\theta(\phi),T)
	+
	\Sigma(x_{-D+1}^0,x_1^n,T)
	-\log\pi_D(T;\beta)
	\Bigg\}\\
=&
\max_{x_1^n\in S_n}
\left\{
\log 
\left(
\max_{T\in\clT(D)}\sup_{\phi\in\Omega(T,m)}
\frac{P(x_1^n|x_{-D+1}^0,\theta(\phi),T)}
{\exp\left\{
	\Sigma(x_{-D+1}^0,x_1^n,T)
-\log\pi_D(T;\beta)
\right\}}
\right)
-\log Q_n(x_1^n)
\right\}\\
=&
\max_{x_1^n\in S_n}
\left[
\log 
\left(
\frac{\mu(x_1^n|x_{-D+1}^0)}
{Q_n(x_1^n)}
\right)
+\log Z(x_{-D+1}^0)
\right]\\
\geq &
	\log Z(x_{-D+1}^0).
\end{align*}
Therefore, in view of~(\ref{eq:partition}),
in order to prove the theorem it suffices to show that
\be
\liminf_{n\to\infty}
\log\left(
\sum_{y_1^n\in A^n}
	\max_{T\in\clT(D)}
	\left[P(y_1^n|x_{-D+1}^0,T)
	\pi_D(T;\beta)\right]
\right)
\geq 0.
\label{eq:targetL}
\ee
To that end we observe that,
replacing the maximum over $T$ by
the expectation with respect to
the posterior of $T$, the
above logarithm is,
\ben
\log\left(
\sum_{y_1^n\in A^n}
	\max_{T\in\clT(D)}
	P(y_1^n,T|x_{-D+1}^0)
\right)
&\geq&
\log\left(
\sum_{y_1^n\in A^n}
	\sum_{T\in\clT(D)}
	\pi(T|y_1^n,x_{-D+1}^0)
	P(y_1^n,T|x_{-D+1}^0)
\right)\\
&\geq&
\sum_{y_1^n\in A^n}
\sum_{T\in\clT(D)}
	P(y_1^n,T|x_{-D+1}^0)
\log
	\pi(T|y_1^n,x_{-D+1}^0),
\een
where the second inequality 
follows from Jensen's inequality.
But the last term above can
be seen to equal the negative
of the conditional
entropy $-H(T|Y_1^n,x_{-D+1}^0)$,
where we recall that, for 
three discrete random variables
$X,Y$ and $Z$, the conditional
entropy of $X$ given $Y$ and $Z=z$
is defined, in the obvious notation,
as,
$$H(X|Y,Z=z)=-\sum_{x,y}P_{X,Y|Z}(x,y|z)\log P_{X|Y,Z}(x|y,z).$$
Now, since the MAP model $T^*(n)$ is
a function of $Y_1^n,x_{-D+1}^0$, by the
data processing property of conditional
entropy~\cite{cover:book},
$$H(T|Y_1^n,x_{-D+1}^0)=
H(T|T^*(n),Y_1^n,x_{-D+1}^0)\leq
H(T|T^*(n),x_{-D+1}^0),$$
where the inequality follows from the
fact that conditioning reduces the entropy~\cite{cover:book}.
Now let $P_{e,n}$ denote the probability
$\BBP(T^*(n)\neq T|x_{-D+1}^0)$, and note
that it tends to zero by Theorem~\ref{thm:modelC}
and dominated convergence. Then, by Fano's inequality~\cite{cover:book},
$$H(T|Y_1^n,x_{-D+1}^0)\leq 
H(T|T^*(n),x_{-D+1}^0)\leq 
h(P_{e,n})+P_{e,n}\log|\clT(D)|,$$
where $h(p):=-p\log p-(1-p)\log(1-p),$ $p\in(0,1)$, denotes the
binary entropy function. And letting $n\to\infty$ we have that,
\ben
\liminf_{n\to\infty}
\log\left(
\sum_{y_1^n\in A^n}
	\max_{T\in\clT(D)}
	P(y_1^n,T|x_{-D+1}^0)
\right)
&\geq&
\liminf_{n\to\infty}
\big[-H(T|Y_1^n,x_{-D+1}^0)\big]\\
&\geq&
\liminf_{n\to\infty}
\big[-h(P_{e,n})-P_{e,n}\log|\clT(D)|\big]
\;=\;0,
\een
establishing~(\ref{eq:targetL}) and completing the proof.
\qed

%
%


\section*{Acknowledgements}

The author wishes to thank Ioannis Papageorgiou for his useful
comments on an earlier draft of this paper.


\bibliographystyle{plain}

\def\cprime{$'$}


\end{document}